\documentclass[useAMS,usenatbib]{mn2e}

\usepackage{graphicx}
\usepackage{color}
\usepackage{amssymb}
\usepackage{arydshln}
\usepackage{amsmath}

\title[Dwarf galaxy merger trees]
  {Numerical simulations of dwarf galaxy merger trees}
\author[A. Cloet-Osselaer et al.]
  {A.~Cloet-Osselaer$^1$\thanks{E-mail:
  Annelies.Cloet-Osselaer@UGent.be}, S.~De Rijcke$^1$, B.~Vandenbroucke$^1$,
  J.~Schroyen$^1$, M.~Koleva$^1$, \and R.~Verbeke$^1$
  \\ $^1$Sterrenkundig Observatorium,
  Ghent University, Krijgslaan 281, S9, 9000 Gent, Belgium }
\date{Accepted , Received ; in original form}

\pagerange{\pageref{firstpage}--\pageref{lastpage}} \pubyear{2014}

\def\LaTeX{L\kern-.36em\raise.3ex\hbox{a}\kern-.15em
    T\kern-.1667em\lower.7ex\hbox{E}\kern-.125emX}

\begin{document}

\label{firstpage}

\maketitle

\begin{abstract}
We investigate the evolution of dwarf galaxies using $N$-body/SPH
simulations that incorporate their formation histories through merger
trees constructed using the extended Press-Schechter formalism. The
simulations are computationally cheap and have high spatial
resolution. We compare the properties of galaxies with equal final
mass but with different merger histories with each other and with
those of observed dwarf spheroidals and irregulars.

We show that the merger history influences many observable dwarf
galaxy properties. We identify two extreme cases that make this
influence stand out most clearly:~(i) merger trees with one massive
progenitor that grows through relatively few mergers and (ii) merger
trees with many small progenitors that merge only quite late. At a
fixed halo mass, a type (i) tree tends to produce galaxies with larger
stellar masses, larger half-light radii, lower central surface
brightness, and since fewer potentially angular momentum cancelling
mergers are required to build up the final galaxy, a higher specific
angular momentum, compared with a type (ii) tree.

We do not perform full-fledged cosmological simulations and therefore
cannot hope to reproduce all observed properties of dwarf
galaxies. However, we show that the simulated dwarfs are not unsimilar
to real ones.

\end{abstract}


\section{Introduction}
Numerical simulations of individual dwarf galaxies have several
advantages over full-fledged cosmological simulations:~one can achieve
very high spatial resolution and one has full control over the initial
conditions, provided the latter are sufficiently realistic and
cosmologically motivated. Thus, it can be easier to study the impact
of certain physical parameters, such as mass or angular momentum, on
the evolution of a galaxy. However, real galaxies obviously do not
evolve in isolation from the rest of the Universe. For one, according
to current cosmological theory, even dwarf galaxies have formed
through a series of mergers in a bottom-up fashion.

Isolated dwarf galaxy simulations are not computationally demanding,
have a well determined initial set-up, and can achieve high spatial
resolution. They can be extended to also include ram-pressure
stripping or interactions with a massive neighbor
\citep{mayer01}. Such simulations have shown that the total galaxy
mass is the main parameter determining the appearance and evolution of
dwarf galaxies \citep{valcke08, revaz09, sawala10}. \cite{schroyen11}
suggest angular momentum as a crucial second parameter that
determines individual star formation modes and offers an
explanation for the observed metallicity gradients \citep{tolstoy04,
  koleva09, tolstoy09, kirby11_3, battaglia11}.

Supposedly more realistic simulations of dwarf galaxies can be
obtained from large ab initio cosmological simulations. However, due
to their low mass, the dwarf galaxies in such simulations are often
seriously undersampled making it difficult to produce robust
predictions for their observational properties \citep{sawala11}. For
example, the dark matter particle mass in the Millennium simulation
\citep{springel05a} and the Millennium-II simulations \citep{boylan09}
is, respectively, 8.6$\times$10$^8$ h$^{-1}$ M$_{\odot}$ and
6.88$\times$10$^6$ h$^{-1}$ M$_{\odot}$. Similarly the Bolshoi
simulations \citep{klypin11} have a mass resolution of
1.35$\times$10$^8$ h$^{-1}$ M$_{\odot}$. Given that dwarfs with masses
similar to the Local Groups dwarfs contain only 10 to 100 dark matter
particles, and their progenitors even less, their merger histories
will be very poorly described by these simulations. Also, the
identification of haloes in large cosmological simulations is not
straightforward and the resulting merger trees can be different
depending on which halo finder method is used \citep{srisawat13}.
Moreover, one has no handle on the number or the properties
(e.g. final mass) of the formed dwarfs. A significant improvement over
this is to re-simulate a small part of a cosmological simulation box
to follow the formation and evolution of a dwarf galaxy of interest in
full detail \citep{governato10, gonzalez13}. Unfortunately, this
requires two simulations to be run and is therefore a computationally
expensive endeavor.

In this paper, we present a third way of producing more cosmologically
sound dwarf galaxy simulations. The Press-Schechter (PS) formalism
\citep{press74} uses the spherical collapse model, which is a simple
model for the nonlinear structure formation in the Universe, to derive
the conditional mass function. The extended Press-Schechter (EPS)
theory \citep{bond91, lacey93} or excursion set approach uses this
conditional mass function to estimate the rate at which smaller
objects merge into larger objects or the halo formation
distribution. With the help of Monte Carlo algorithms a merger tree
can be constructed in a top-down fashion starting from its final mass.
There are many algorithms available to investigate structure formation
based on this method. A detailed comparison of existing Monte Carlo
algorithms and a general overview of the EPS theory can be found in
\cite{zhang08}, along with a comparison of the algorithms of
\cite{kauffmann93, lacey93, somerville99, cole00} and three new
algorithms. However, the results of the algorithms that use EPS
overpredict the abundance of small haloes and underpredict the
abundance of larger haloes with increasing redshift compared to the
result of the cosmological $N$-body simulations \citep{lacey94,
  tormen98, sheth99, zhang08}. This is likely due to the ``spherical''
approximation which is used in EPS while real haloes are rather
triaxial \citep{bardeen86}. But, as the results from the spherical
collapse model produce merger trees with statistical properties which
have the same trends with mass and redshift as merger trees from the
Millennium simulation, \cite{parkinson08} adapted the {\sc GALFORM}
algorithm of \cite{cole00} to fit the conditional mass function of the
Millennium simulation.

We propose to use EPS to produce a merger tree that fixes the timing
of the mergers leading up to a galaxy of a given mass at $z=0$. The
orbital parameters of the individual mergers are sampled from
probability distribution functions derived from cosmological
simulations. We then use an $N$-body/SPH code to simulate in full
detail the merger sequence and the build-up and evolution of the
galaxy. Using this approach, one is able to build simulated galaxies
in a more cosmologically realistic way while retaining some of the
benefits of the isolated-galaxy simulations, such as high resolution
and control over the final galaxy mass.

In section \ref{section:numerical_details}, we provide more details
about the numerical methods in the code that was used to simulate
dwarf galaxies. An analysis of the simulations is given in section
\ref{section:analysis}, where our models are compared with
observations in terms of their location on the observed kinematic and
photometric scaling relations. In section \ref{section:results} we
discuss the obtained results and we formulate our conclusions in
section \ref{section:conclusion}.


\section{Simulations}
\label{section:numerical_details}
For the simulations we use a modified version of the $N$-body/SPH code
{\sc Gadget-2} \citep{springel05}. This code is extended with star
formation, feedback and metal dependent radiative cooling. In the
past, this code has been used by our group to simulate isolated dwarf
galaxy models with cosmologically motivated initial conditions,
meaning a NFW halo \citep{nfw96} for the dark matter, a
baryon fraction in agreement with the cosmological model, etc.  The
results of these simulations are discussed in \cite{valcke08},
\cite{valcke10}, \cite{schroyen11}, \cite{cloetosselaer12}, and
\cite{schroyen13}.  From these studies we can conclude that the
isolated models are in agreement with the kinematic and photometric
scaling relations of observed dwarf galaxies and the results of other
simulations \citep{revaz09, sawala10}.

We build on our experience with isolated models in order to construct
a dwarf galaxy with a hierarchical structure formation history. First,
we construct a merger tree (see paragraph
\ref{subsection:mergerTrees}) whose leaves are populated with isolated
dwarf galaxy models with cosmologically motivated initial conditions
(ICs) (see paragraph \ref{subsection:IC}). These protogalaxies are
then evolved and merged using the $N$-body/SPH code (see paragraph
\ref{subsection:code}). 

As for our isolated simulations we do not aim to specifically simulate
either late-type or early-type dwarfs. The classification of a galaxy
would in any case depend on the presence of gas-removing processes,
such as ram-pressure stripping, and on when exactly during the
star-formation duty cycle the galaxy is observed, as explained in
e.g. \cite{schroyen13}. Given the absence of external gas-removing
processes, our simulated galaxies keep their gas until the end and
could therefore be classified as late-types. Before this gas removal,
there is little reason to suggest that late and early types had a
different evolution.

For the visualization and
analysis of our results we use our own public available software
package {\sc HYPLOT}. This software is freely available from
SourceForge\footnote{http://sourceforge.net/projects/hyplot/} and is
used for all the figures in this paper.

\subsection{Initial conditions isolated galaxies}\label{subsection:IC}
We briefly describe the initial setup of the isolated models as each
of the members of the merger tree will be setup in isolation.  For the
dark matter halo an NFW profile is used with a density
profile:
\begin{equation}
 \rho_{\rm NFW}(r) = \frac{\rho_{\rm s}}{(r/r_{\rm s})(1+r/r_{\rm
     s})^{2}}
\end{equation}
where $\rho_{\rm s}$ and $r_{\rm s}$ are respectively the
characteristic density and the scale radius. The \cite{strigari07}
relations for the concentration parameter of dwarf dark matter haloes
are used for the determination of $\rho_{\rm s}$ and $r_{\rm s}$:
\begin{eqnarray}
 c &\approx& 33 \bigg(\frac{M_{\rm h}}{10^8 M_{\odot}}\bigg)^{-0.06}
 \\ \rightarrow& \rho_{\rm s} =&
 \frac{101}{3}\ \frac{c^{3}}{\ln(1+c)-c/(1+c)}\ \rho_{c}
 \\ \rightarrow& r_{\rm s} =& \bigg(\frac{M_{\rm h}}{4\pi
   \rho_s}\frac{1}{\ln(1+c)-c/(1+c)}\bigg)^{1/3}
\end{eqnarray}
where $\rho_{c}$ is the critical density of the universe at $z=0$ and
M$_{h}$ is the halo mass in units of solar masses. The concentration
parameter $c$ is defined as the ratio of the virial radius, r$_{\rm
  max}$, at which the dark matter density profile is cut off, to the
scale radius, $r_s$. For the gas cloud, we use a pseudo-isothermal
density profile. For the detailed implementation of this gas halo we
refer to \cite{schroyen13}. The initial gas metallicity is set to
$10^{-4}\ Z_{\odot}$, the initial temperature is $10^4$ K. We use a
gravitational softening length of 0.03~kpc for all particles.

We use a flat $\Lambda$-dominated cold dark matter cosmology with $h =
0.71, \Omega_{\rm tot} = 1, \Omega_{\rm m} = 0.2383, \Omega_{\rm DM} =
0.1967$.  The baryonic mass or gas mass in our models is set to be
0.2115 times that of the dark-matter, in accordance to the employed
cosmology.

\subsection{The code}\label{subsection:code}
We use a modified version of the Nbody-SPH code {\sc Gadget-2}
\citep{springel05} which is extended by \cite{valcke08} with star
formation, feedback and radiative metallicity dependent cooling. We go
trough the implementations of these extensions in the following
subsections.

\subsubsection{Star formation}
Star formation happens in dense, cold and gravitationally collapsing
gas regions which are too small to resolve in our simulations, hence,
we are forced to implement star formation as a subgrid formalism. In
our code, we select star formation regions with the following
criteria:
\begin{eqnarray}
 \rho_{\rm g} & \geq & \rho_{\rm SF} \\ T & \leq & T_{\rm c} = 15000
 \mathrm{K} \\ \vec{\nabla}.\vec{v} & \leq & 0.
\end{eqnarray}
The most important star formation criteria is the first one, the
density threshold, for which we use a value of 10 amu cm$^{-3}$. This
is in agreement with the trend to use a high density threshold in
simulations \citep{governato10,guedes11,cloetosselaer12,schroyen13}
which map the regions of active star formation more accurate.  Gas
particles that fulfill these criteria can become star particles
depending on a Schmidt law. This law connects star formation with the
gas density, $\rho_g$, the dynamical time, $t_g$, and the parameter
c$_{\star}$, which is the dimensionless star formation efficiency:
\begin{equation}
\frac{\mathrm{d}\rho_{\rm s}}{\mathrm{d}t} =
-\frac{\mathrm{d}\rho_{\rm g}}{\mathrm{d}t} = c_{\star}\frac{\rho_{\rm
    g}}{t_{\rm g}}. \label{cstar}
\end{equation}
As shown by other authors (e.g. \citet{stinson06,revaz09}) and based
on our own experience, there is a wide range of values for the
c$_{\star}$ parameter that produces dwarf galaxies with acceptable
observable properties and self-regulated star formation. Any value
within this range is equally valid. For the simulations presented here
we use a value of 0.25 for the c$_{\star}$ parameter, the same value
as in \cite{cloetosselaer12}.

By using a high density threshold, star formation occurs more in small
gas clumps and is less centrally concentrated, in agreement with
observed galaxies \citep{schroyen13}.

\subsubsection{Feedback and cooling}
The code is implemented with feedback from Type Ia supernova (SNIa),
Type II supernove (SNII) and stellar winds (SW) as described in
\cite{valcke08}. When stars die they deliver thermal energy to the ISM
and they enrich the gas. Star particles represent single-age
single-metallicity stellar populations (SSP) with a Salpeter
initial-mass function \citep{salpeter55}.  The feedback is released as
thermal feedback to the gas particles close to the star particle,
e.g. within the SPH smoothing radius and according to the smoothing
kernel of the gas particle the star particle originates from. Massive
stars, with low mass limit $m_{SNII, l}=8\ M_{\odot}$ and high mass
limit $m_{SNII, h}=60\ M_{\odot}$, die as SNII supernove, while less
massive stars, with low mass limit $m_{SNIa, l}=3\ M_{\odot}$ and high
mass limit $m_{SNIa, h}=8\ M_{\odot}$, will explode as SNIa
explosions. As massive stars will have a short life, they will release
their feedback quite fast after a star is born. For the SNIa, we
employ a delay of 1.5 Gyr. The total energy released by supernova
explosions is set to 10$^{51}$ ergs and for stellar winds to 10$^{50}$
ergs. This energy is distributed at a constant rate during their main
sequence lifetime. The lifetime is a function of the mass $m$ of the
star and is given by \citep{david90}:
\begin{equation}
\log t(m) = 10-3.42 \log(m)+0.88(\log(m))^2
\end{equation} 
The energy effectively absorbed by the ISM is obtained by multiplying
these energies with the feedback efficiency factor, which is set to
0.7. As discussed in \cite{cloetosselaer12} there is a degeneracy
between the density threshold for star formation and the feedback
efficiency factor and as indicated in the previous section, the SN
efficiency will also influence the self-regulation of the star
formation. From this article, we copied the set of values that has
shown to result in dwarf galaxies with properties comparable to real
dwarf galaxies. In addition, a high density threshold for star
formation in combination with a suitable feedback efficiency partially
reduces the final stellar mass of the galaxy which is generally
overpredicted in simulations \citep{scannapieco12, sawala11}.

For the cooling we use the metallicity-dependent cooling curves from
\cite{sutherland93}, which describe the cooling of gas for different
metallicities down to a temperature of 10$^4$ K. These cooling curves
are extended to lower temperatures using the cooling curves of
\cite{maio07}. A particle which is heated by supernova explosion will
not be able to cool radiatively during the time-step where the SN
explosion occurred, this in order to correct for the fact that our
resolution cannot resolve the hot, low density cavities that are
generated by the supernova explosions.

\begin{table}
\begin{tabular}{rrrr}
\hline 
$\mathrm{M_{\rm halo}}$ & $M_{\rm res}$ & $\mathrm{\#DM\ particles}$ & $m_{\rm p}$ \\ \hline 10$\times$10$^{9}$
M$_{\odot}$ & 0.75$\times$10$^{8}$ M$_{\odot}$ & 800 000 & 12 500
M$_{\odot}$\\ 7.5$\times$10$^{9}$ M$_{\odot}$ & 0.50$\times$10$^{8}$
M$_{\odot}$ & 400 000 & 18 750 M$_{\odot}$\\ 5.0$\times$10$^{9}$
M$_{\odot}$ & 0.25$\times$10$^{8}$ M$_{\odot}$ & 400 000 & 12 500
M$_{\odot}$\\ 2.5$\times$10$^{9}$ M$_{\odot}$ & 0.25$\times$10$^{8}$
M$_{\odot}$ & 400 000 & 6 250 M$_{\odot}$\\ 1.0$\times$10$^{9}$
M$_{\odot}$ & 0.1$\times$10$^{8}$ M$_{\odot}$ & 400 000 & 2 500
M$_{\odot}$\\ \hline
\end{tabular}
\caption{Details of the input parameters for the merger trees. In the
  first column the halo mass is shown, the second column shows the
  mass resolution which is the smallest possible halo mass. The third
  column shows the amount of dark matter particles in the simulation
  and the last column shows the mass of one dark matter
  particle. \label{Mergers_table}}
\end{table}
 
\subsection{Merger trees} \label{subsection:mergerTrees}

\begin{figure*}
\begin{minipage}[ht]{\linewidth}
\begin{center}
 \centering 
 \includegraphics[width=0.95\textwidth,clip]{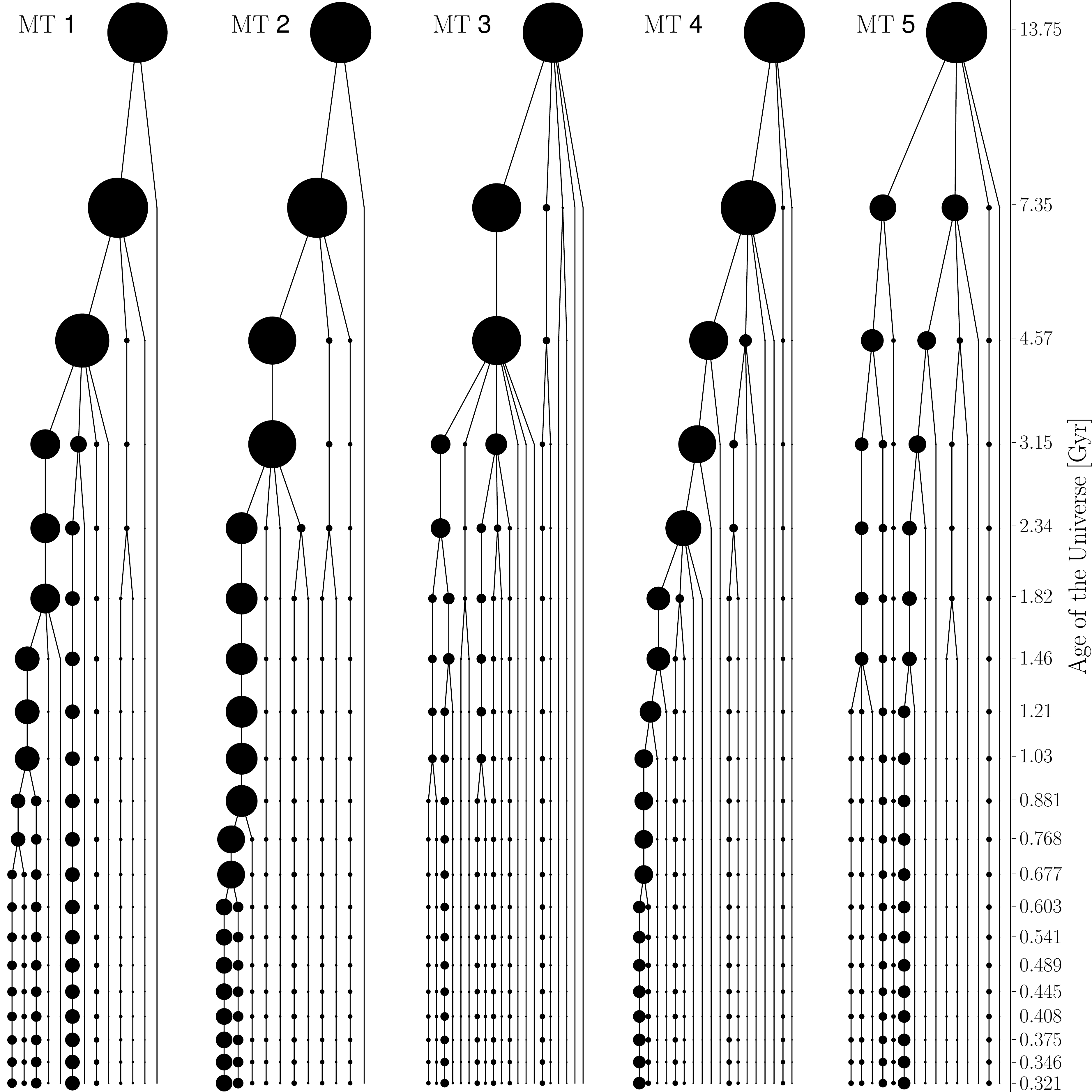}
 \caption{The 5 merger trees of the haloes with a final mass of
   M$_{h}$=2.5$\times$10$^{9}$ M$_{\odot}$. The size of the circles
   gives an indication of the mass of the halo. The evolution is shown
   as a function of the age of the universe which corresponds with the
   redshift range from 13.5, when the Universe is 0.32 Gyr old to
   $z=0$.  \label{fig:mergers_all}}
 \end{center}
 \end{minipage}
\end{figure*}

We have used the {\sc GALFORM} algorithm \citep{cole00}, as modified
by \cite{parkinson08}, to construct merger trees.  This algorithm is
based on the EPS theory which starts from an initial Gaussian random
density fluctuation field and uses the analytical model of
cosmological spherical collapse to construct a density threshold above
which a halo becomes virialized. For a halo of a given mass at a
certain redshift, it predicts the conditional mass function of its
progenitors at a higher redshift. With Monte Carlo techniques a path
can be constructed from the final mass of the galaxy, the root of the
tree, to the leaves, which are the smallest galaxies considered in the
calculation, at some high redshift. \cite{parkinson08} adjusted
the algorithm to fit the conditional mass function of the Millennium
simulation \citep{springel05}. The construction of the merger tree
proceeds from its root at $z=0$ to its leaves, which we place at a
lookback time of approximately 13.5 Gyr, corresponding to a redshift
of $z$=13.5. This redshift interval is divided into 20 bins of equal
size. A few examples of the merger trees that are used can be found in
Fig. \ref{fig:mergers_all}. There, the sizes of the circles give an
indication of the mass of the haloes.

As we do not run cosmological simulations, our haloes do not grow in
time due to accretion. As visible in the merger trees in
Fig. \ref{fig:mergers_all}, the only way to gain mass is by
merging. However, the output of the merger tree algorithm takes mass
accretion into account. When the mass growth in a timestep is smaller
than some resolution mass it will be considered accreted mass and it
will be added to the main halo mass. As we want our final mass to be
well determined we distribute the accreted mass of a parent halo over
its progenitors in proportion to their masses. This way, the entire
final mass is already present in the simulations from a redshift of
$z=13.5$, but it is distributed over all the haloes. An overview of
our different merger tree simulations can be found in Table
\ref{Mergers_table}. This table shows the final masses of the haloes,
their resolution mass which is used in the merger tree algorithm, the
number of DM particles which is used to simulate the dark matter halo
and the mass resolution of the dark matter in the simulation. Thus,
the tree construction process provides us with the masses of the
leaves of the merger tree in combination with their future merger
history. We now have to place these leaf galaxies on suitable orbits
in order to merge at the appropriate time.

For each merger event, we select the most massive dark-matter halo
among the haloes that need to be merged. This we call the ``primary''
halo. In case of a binary merger there is only one ``secondary'' halo;
in case of a multiple merger, there can be several secondaries. Each
primary/secondary merger is treated as a 2-body problem. We equate the
time of the merger provided by the merger tree with the pericenter
passage of the primary/secondary couple. First, we calculate the
virial velocity of the primary halo. Then, we use the 2D probability
distribution function of \cite{benson05} to randomly draw a value for
the radial and tangential velocities of the incoming secondary halo as
it crosses the primary's virial radius, denoted by $v_{r}$ and
$v_{\theta}$, expressed in units of the primary's virial velocity:
\begin{equation}
 f(v_{r}, v_{\theta}) = a_{1} v_{\theta}
 \exp{\{-a_{2}(v_{\theta}-a_{9})^{2}-b_{1}(v_{\theta})[v_{r}-b_{2}(v_{\theta})^2]\}}
\end{equation}
with
\begin{eqnarray}
 b_{1}(v_{\theta}) = a_{3} \exp{[-a_{4}(v_{\theta}-a_{5}^2)]}
 \\ b_{2}(v_{\theta}) = a_{6} \exp{[-a_{7}(v_{\theta}-a_{8}^2)]}
\end{eqnarray}
and we used for the values a$_{1-9}$ respectively
\begin{center}
\resizebox{\columnwidth}{!}{
\begin{tabular}{c|c|c|c|c|c|c|c|c}
 a$_{1}$ & a$_{2}$ & a$_{3}$ & a$_{4}$ & a$_{5}$ & a$_{6}$ & a$_{7}$ &
 a$_{8}$ & a$_{9}$ \\ \hline 6.38 & 2.30 & 18.8 & 0.506 & -0.0934 &
 1.05 & 0.267 & -0.154 & 0.157
\end{tabular}
}\end{center} In practice, almost all trajectories are nearly
parabolic with an ellipticity close to 1. As \cite{benson05} finds
only a very weak correlation between the spatial distribution of
subsequent mergers, we draw the orbital plane positions from an
isotropic distribution. From this velocity information at the virial
radius, we determine the orbital elements of the corresponding Kepler
orbit. We want to follow each merger starting 2~Gyr before its
pericenter passage. Therefore, we introduce the secondary halo into
the simulation at a position and with a velocity that would bring it
to the pericenter of its Kepler orbit 2~Gyr in the future. Obviously,
since galaxies are deformable, they will not adhere to these Kepler
orbits. This with the exception of the mergers occurring during the
first 2 Gyr of a simulation. In that case, the time to reach
pericenter is set to be the difference between the merging time and
the start of the simulation. We note that it is perfectly possible for
a merger to start when the previous merger that formed the primary is
still ongoing, leading to complex multi-galaxy encounters.

\begin{figure}
 \includegraphics[width=0.5\textwidth,clip]{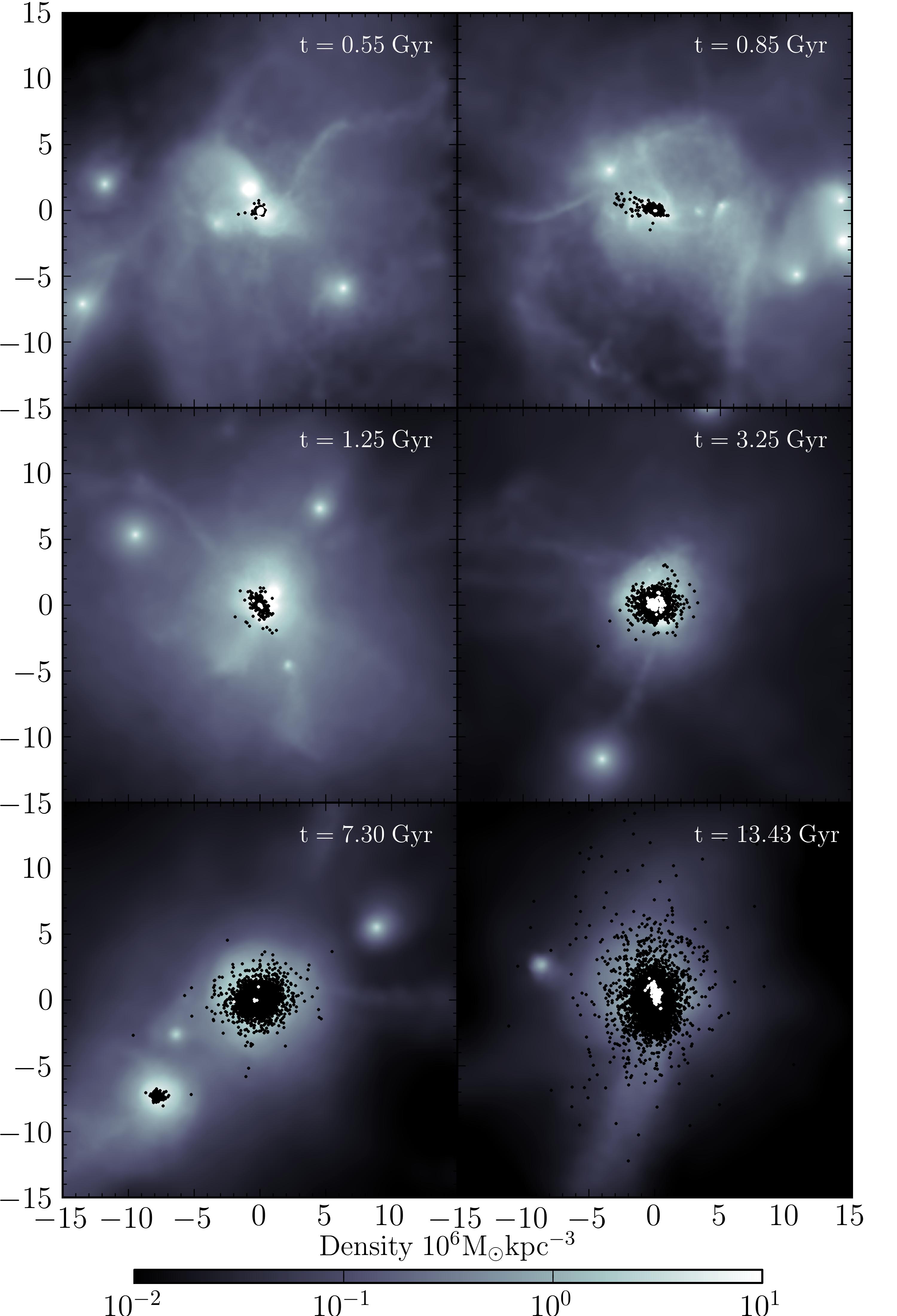}
 \caption{Snapshots of the merger simulation corresponding with tree
   MT4 from Fig. \ref{fig:mergers_all}. The grayscale represents the
   gas density, the white dots show the young stars which are younger
   then 0.1 Gyr and the black dots show all the stars in the
   galaxy. In each panel the snapshot time is indicated, the $x$ and
   $y$-labels are in kpc.\label{fig:nicepic}}
\end{figure}

In Fig. \ref{fig:nicepic}, a few snapshots of the evolution of a
typical merger simulation are shown. The gas density is rendered in
grayscale, the young stars ($\leq$0.1 Gyr) are plotted as white dots,
the other star particles as black dots. The most lightweight galaxies
cannot compress the gas to densities above the star-formation
threshold and remain starless. In fact, initially only the most
massive halo is able to ignite star formation (see first panel). The
addition of smaller galaxies with gas can trigger bursts of enhanced
star formation. With time, the merger activity subsides (see last
panel).

\section{Analysis}
\label{section:analysis}

\subsection{Star formation histories}
\label{section_SFH}
\begin{figure}
 \centering
 \includegraphics[width=0.5\textwidth]{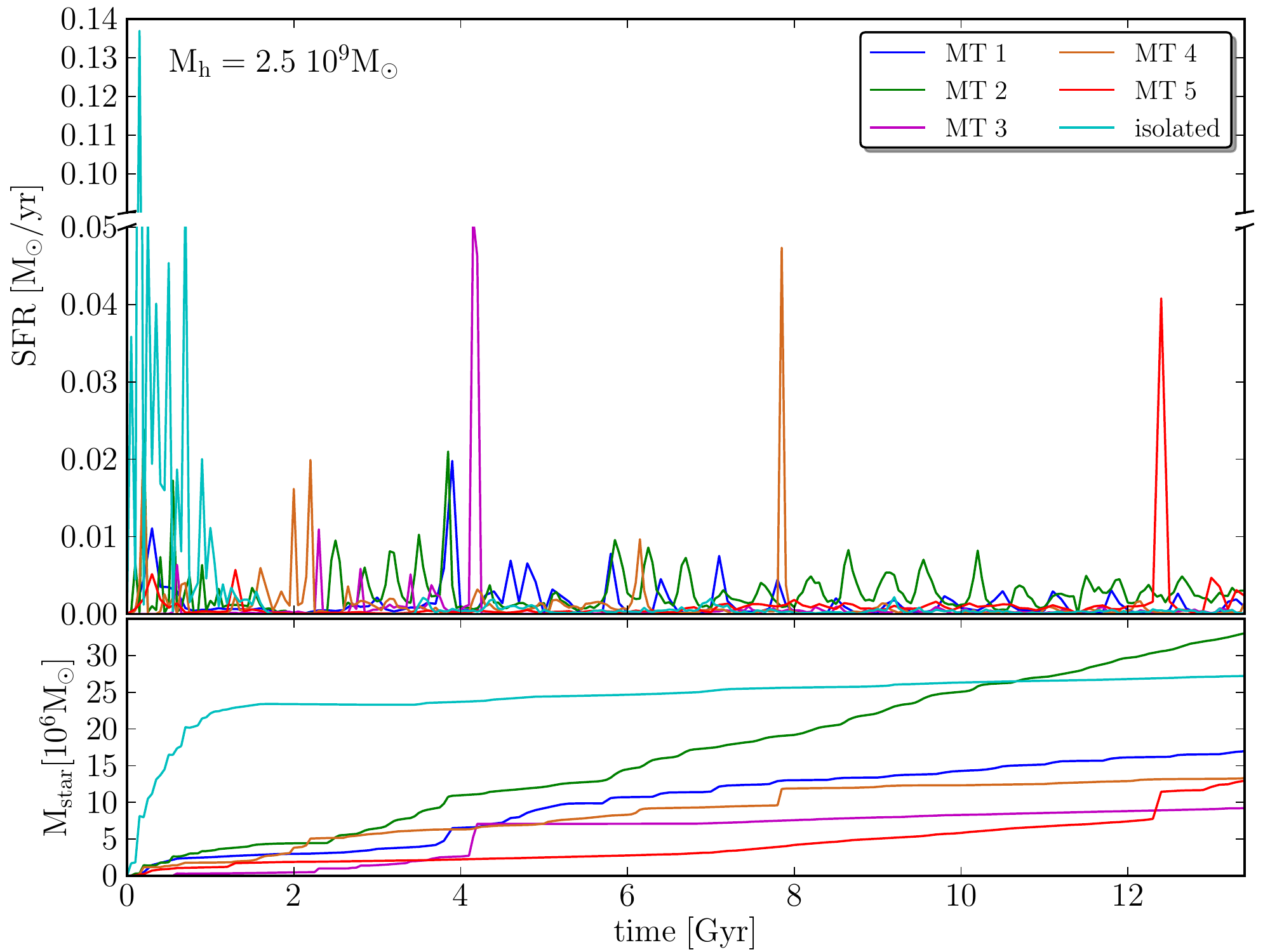}
 \caption{Top panel:~the total SFR of several merger trees and a
   reference isolated simulation with the same final halo mass of
   2.5$\times$10$^{9}$ M$_{\odot}$ as a function of time. Bottom
   panel:~the stellar mass as a function of time.\label{fig:SFR} (A
   color version of this figure is available in the online journal.)}
\end{figure}

\begin{figure}
 \centering
 \includegraphics[width=0.5\textwidth]{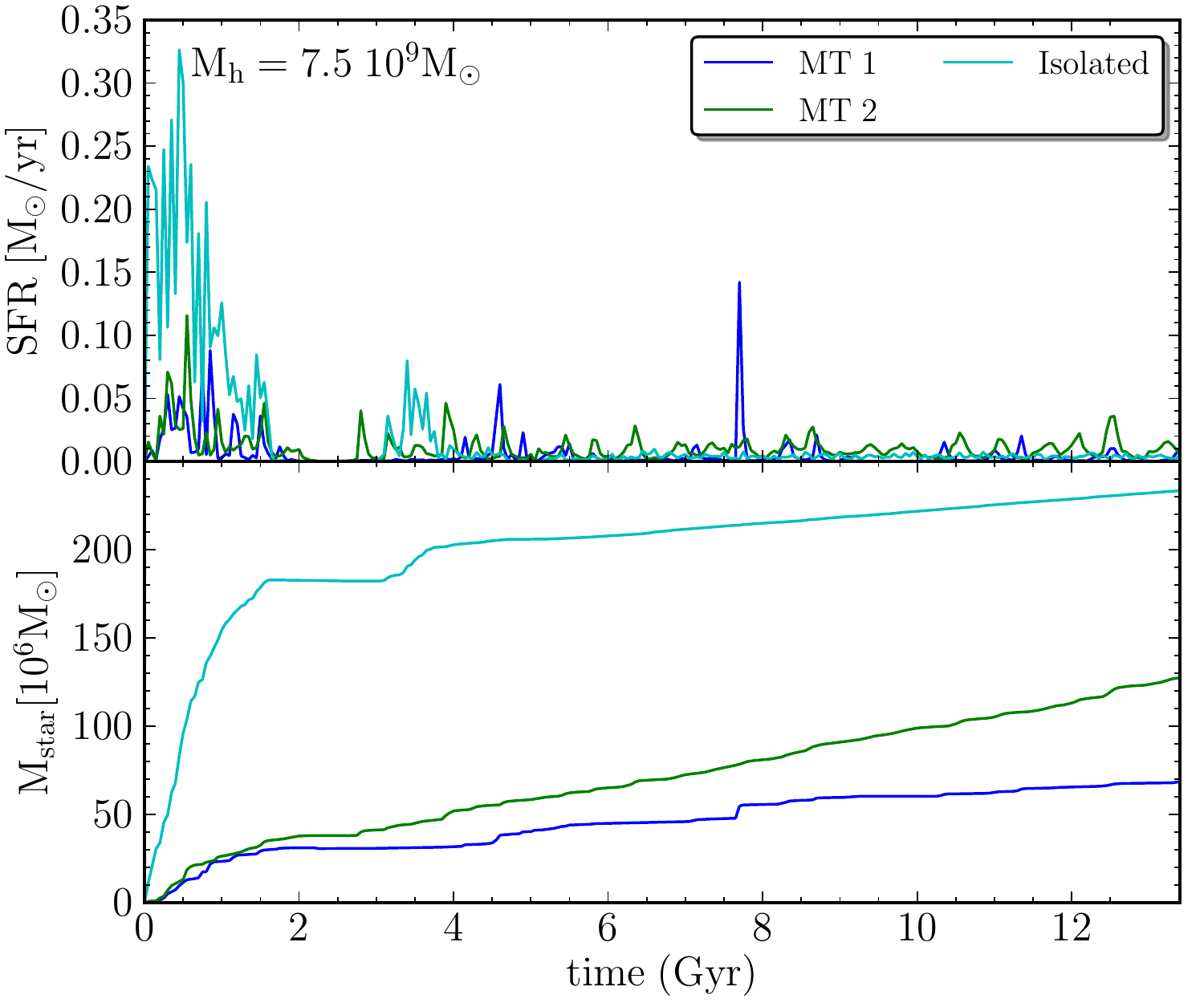}
 \caption{Identical as Fig. \ref{fig:SFR} but for a final halo mass of
   7.5$\times$10$^{9}$ M$_{\odot}$. (A color version of this figure is
   available in the online journal.)\label{fig:SFR2}}
\end{figure}

In Fig. \ref{fig:SFR} and Fig. \ref{fig:SFR2} the star-formation
histories (SFHs) of different merger trees are shown, all with the
same final halo mass of respectively 2.5$\times$10$^{9}$ M$_{\odot}$
(Fig. \ref{fig:SFR}) and 7.5$\times$10$^{9}$ M$_{\odot}$
(Fig. \ref{fig:SFR2}). For comparison, the SFH of an isolated model
with the same final halo mass is plotted in these figures.

\subsubsection{Isolated galaxies}\label{subsubsection:isolated}

The isolated models form most of their stellar mass during the first 2
Gyr of the simulation after which star formation shuts down for
approximately 2 Gyr due to the depletion of gas by SN feedback which
prevents the gas density to reach the threshold for star
formation. The star formation restarts around 4 Gyr when part of the
gas has been able to return. After that, star formation can proceed in
an episodic fashion, such as in the isolated model presented in
Fig. \ref{fig:SFR}, or as low-level residual star formation, as
in the isolated model presented in Fig. \ref{fig:SFR2}. The major
difference between the different mass models in isolation is the
amplitude of the star formation, which increases with increasing mass.

\subsubsection{Merged galaxies}\label{subsection:merged}

In Figs. \ref{fig:SFR} and \ref{fig:SFR2}, the sum of the SFR of all
the members of a merger tree is plotted. In the first two Gyrs of the
simulation, it is difficult to disentangle the difference between star
formation occurring naturally inside any given halo and star formation
prompted by mergers. After this period, the peaks in the SFH can be
shown to correspond to the major merger events in the merger trees.
For example, in Fig. \ref{fig:SFR}, we see a peak for MT3 around 4.25
Gyr in the simulations which corresponds with the major merger in MT3
around 4.57 Gyr in Fig. \ref{fig:mergers_all}. Similarly, a peak in
the SFH of MT4 is seen around 7.8 Gyr and in MT5 around 12.5 Gyr,
these correspond respectively with a major merger around 7.35 Gyr for
MT4 and around 13.75 Gyr for MT5.  We see that these SF peaks are in
close agreement with the desired merger history we wanted to generate.
However, small deviations are possible and they are due to:
\begin{enumerate}
 \item The 2-body approximation that is used to determine the orbital
   parameters. In the simulation multiple mergers will be possible.
 \item The calculated merging time is the time to pericenter but the
   large peaks in the star formation will occur only when they actually
   merge.
\end{enumerate}
As in binary merger studies \citep{dimatteo07, cox06,torrey12,
  scudder12}, we see that the peaks in the SFR occur at the end of the
merging process of two galaxies. At the first pericenter passage their
is a modest increase of the SFR due to tidal squeezing of the gas
while a larger increase is noticeable when the galaxies really
collide. For example in Fig. \ref{fig:SFR}, showing the results for
models with a final halo mass of M$_{h}$=2.5 10$^9$ M$_{\odot}$, the
small SF peaks at 3.65 Gyr of MT3 and at 6.2 Gyr at MT4 are created by
the first pericenter passage of the halo before the large peak in
SFR. There is no such first small peak for MT5 since the merger
proceeds very rapidly.  However, now the main star-formation peak is
somewhat broadened. MT2, and to a lesser extent also MT1, lacks strong
starbursts that would otherwise suppress subsequent star
formation. Its minor mergers keep stirring up the gas and cause many
small star-formation events. Its mass therefore gradually builds up
and, in the case of MT2, eventually exceeds that of the isolated
model.

We also see a double peak in the SFH for the more massive merger tree
MT1 with final halo mass of M$_{h}$=7.5 10$^9$ M$_{\odot}$ in
Fig. \ref{fig:SFR2} at 7.2 Gyr and 7.7 Gyr and a triple peak due to a
merger with three components at 4.15 Gyr, 4.6 Gyr and 4.8 Gyr. MT2 in
Fig. \ref{fig:SFR2} has some major mergers very early on in the
simulation but for the main part of its evolution it is has a
continuously supply of gas by minor mergers.

\cite{dimatteo07} did a statistical study of binary interactions and
mergers of galaxies of all morphologies from ellipticals to late type
spirals. However, our sample is less numerous and contains less
massive and less disky systems but we can check if we observe similar
trends. To start with, a negative correlation was found for the peak
star-formation rate and the strength of the tidal interaction between
a galaxy pair at first pericenter passage. The latter can be
quantified by the pericenter distance, $r_p$, the pericenter velocity,
$v_p$, or the tidal parameter, $T_p$, defined as the sum of the tidal
forces of each component each described by:
\begin{equation}
  T_{p,i} = \log_{10}\Big[
    \frac{M_{comp}}{M_i}\Big(\frac{D_i}{r_p}\Big)^3\Big] \ \ i=1,2
\end{equation}
with $M_i$ the mass of the galaxy, $M_{comp}$ the mass of the
companion, $r_p$ the pericenter distance and $D_i$ the scalelength of
the galaxy, calculated as the radius containing 75\% of the dark
matter mass. Although we of course have much less data to rely on,
Table \ref{SFRpeak} shows a similar trend for the most massive major
mergers of MT3, MT4 and MT5 of the models with final halo mass of
M$_{h}$=2.5 10$^{9}$ M$_{\odot}$, where we see that a decrease in
pericenter distance corresponds to an increase in the amplitude of the
SF peak. Similarly, an increase in the velocity at the pericenter,
$v_p$, corresponds in the models with a decrease in SFR$_{peak}$ in
our data, although \cite{dimatteo07} found no correlation between
these two parameters. However, our models have a similar trend as the
models of \cite{dimatteo07} where an increase of SFR$_{peak}$ occurs
when the characteristic encounter time, $t_{enc}=\frac{r_p}{v_p}$
increases. The explanation for these trends provided by
\cite{dimatteo07} is that a gentler first pericenter passage allows
the orbiting galaxies to retain more of their gas for future
consumption during the final merger phase.

\cite{dimatteo07} found a clear trend for galaxy pairs to have lower
peak star formation rates when they merge if they experienced intense
tidal forces at first pericenter passage. This is due to two
effects. On the one hand, stronger tidal squeezing on the way to
pericenter leads to a slightly enhanced gas consumption by star
formation around pericenter passage while, on the other hand, stronger
expansion of the outer parts of the system after pericenter passage
induces a more significant loss of gas in tidal tails. We find a
similar trend that galaxies which endure a strong tidal force during
their first pericenter passage have lower SF peaks when they merge.

\begin{table}
\begin{tabular}{llllll}
\hline MT & $r_p$ & $v_p$ & $t_{enc}$ & SFR$_{peak}$ & $T_p$\\ & [kpc]
& [km/sec] & [Myr] & [M$_{\odot}$/yr] & \\ \hline \hline MT3 & 2.36
kpc & 40.5 & 56.9 & 0.051 & 5.40 \\ MT4 & 0.91 kpc & 141.3 & 6.3 &
0.049 & 6.71 \\ MT5 & 0.12 kpc & 398.9 & 0.29 & 0.040 & 9.53 \\ \hline
\end{tabular}
\caption{The properties of the most massive major mergers of MT3, MT4
  and MT5 of the models with M$_{h}$=2.5 10$^{9}$ M$_{\odot}$. The
columns show (1) the pericenter distance, (2) the velocity at the
pericenter, (3) the duration of the encounter, (4) the peak in the SF
due to the merger, and (5) the tidal parameter.\label{SFRpeak}}
\end{table}

Table \ref{table_finalprop} shows the final stellar mass of the
different simulations. For the same halo mass, the merger simulations
produce less stars than the isolated simulations. So, while mass is
the main parameter determining the properties of isolated simulated
galaxies, this is no longer true for the merger simulations. Mergers
are able to fling large amounts of gas to large radii, where it is
inaccessible for star formation and the merger history will determine
when gas will be delivered to the center of a galaxy.

In addition, we can distinguish two extreme types of merger trees
which reflect most clearly how the merger history influences a
galaxy's star formation history and final stellar mass. On the one
hand, merger trees can have a massive progenitor present early on in
the simulation which subsequently grows through minor mergers, such as
MT1, MT2, and, to a lesser extent, MT4 (see
Fig. \ref{fig:mergers_all}). At the other extreme, there are merger
trees with many low-mass progenitors that merge only late in cosmic
history, such as MT3 and MT5. In the former, the massive progenitor is
already relatively efficient at forming stars from the start of the
simulation while subsequent minor mergers will fuel further star
formation. This leads to a continuously increasing stellar mass. In
the latter, the many low-mass progenitors are inefficient star formers
and the stellar mass increases mostly during merger-induced
bursts. The former type of merger tree also leads to galaxies with
higher stellar masses at a given halo mass than the latter type. Of
course, merger trees fill in the continuum between these two extreme
types. For instance, MT4 has a merger tree that is intermediate
between the two extreme cases.

Recently, observed SFHs, derived from color-magnitude diagrams, have
become available for sizable samples of dwarf galaxies
\citep{monelli10a, monelli10b, cole07b, weisz11, hidalgo11, mcquinn10,
  mcquinn10b}. The time resolution of these SFHs can be as good as 10
Myr for the most recent epochs, deteriorating to over 500 Myr for
stellar populations older than 1 Gyr. The SFHs derived by
e.g. \citet{mcquinn10} for the last 1.5~Gyr are well resolved and show
that the SFRs of dwarf galaxies can fluctuate strongly and
erratically, with no discernible periodicity. These authors find that
a burst produces between 3 and 26~\% of the final stellar mass in the
observed dwarfs. In our simulations, a starburst produces between 7~\%
and 29~\% of the final stellar mass. A notable exception is the
extreme, 6-fold merger in simulation MT3 with halo mass
2.5$\times$10$^{9}$ M$_{\odot}$ at 4 Gyr, which produces 47~\% of the
stellar mass. The observed bursty dwarfs have SFRs that vary from
0.0003~M$_{\odot}$/year (Antlia) over 0.05~M$_{\odot}$/year (IC4682)
up to 0.4~M$_{\odot}$/year (NGC5253). Our simulated dwarfs have mean
SFRs of the order 0.005-0.01~M$_{\odot}$/year, comparable to
e.g. UGC4483, NGC4163, UGC6458, and NGC6822. The observed ratio of the
burst peak SFR to the mean SFR falls in the range $b \sim 3 - 14$. The
simulated galaxies see peak increases of the order of $b\sim 10$ in
the strongest starbursts and $b\sim 2$ in the weakest bursts.

In section \ref{subsubsection:isolated} we already pointed out that
the isolated non-rotating galaxy simulations often have episodic
SFHs. The merger tree SFHs, on the other hand, are much more erratic
and variable, without fixed periodicity, much like the SFHs of real
dwarfs.

\subsection{Scaling relations}
\label{section_scaling_relations}

Here we compare the gross properties of the simulated dwarfs, both
with and without merger trees, with observed kinematic and photometric
scaling relations at $z=0$. For the observational data: the data from
\cite{derijcke05} are converted from B-band to V-band using the
relation: (B-V)=0.7.  From the \cite{graham03} paper we converted
their B-band magnitudes into V-band magnitudes as described in
\cite{derijcke09} (using a B-V colour-magnitude relation constructed
from the M$_{V}$-(V-I)-[Fe/H] relation in combination with SSP models
for 10-Gyr-old stellar populations (from \cite{vazdekis96})).  The LG
data and the data from \cite{derijcke09}, \cite{grebel03},
\cite{hunter06}, \cite{dunn10} and \cite{kirby13} are presented in the
V-band, so no transformation was needed. The V-band magnitudes of
\cite{vanzee00} and \cite{vanzee04} are deduced from their B-band
magnitude and their B-V color. For the Antlia data from Smith Castelli
2008, the C-T$_{1}$ colours were (as in \cite{derijcke09}) converted
to the C-T1 colours using empirical (C-T$_{1}$)-[Fe/H] and
[Fe/H]-(V-I)-relations. For the $\sigma$-L$_{B}$ plot we need the data
in B-band, which was available for all the data with the exception of
\cite{geha03} where we transform the V-band magnitude to the B-band
magnitude using M$_{V}$=M$_{B}$-0.7. We present the dIrr, dSph and dE
data in Fig. \ref{fig:bigFig} using respectively black (magenta),
light-grey (yellow) and dark-grey (green) dots.

\subsubsection{Half-light radius R$_{e}$}\label{subsection:Re}

\begin{figure*}
\begin{minipage}[ht]{\linewidth}
\begin{center}
 \centering 
 \includegraphics[width=0.85\textwidth,clip]{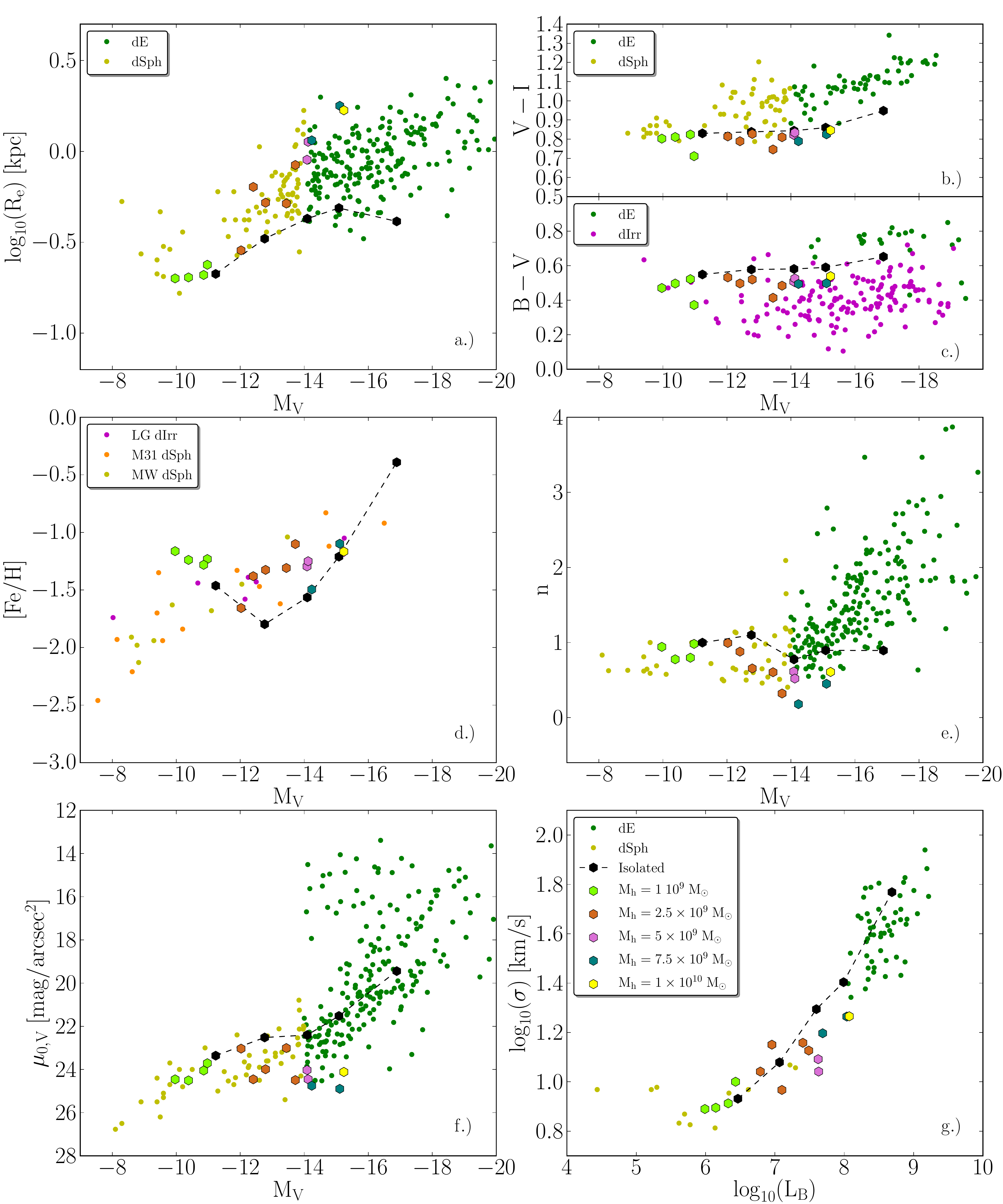}
 \caption{Some scaling relations and the surface brightness parameters
   as a function of the magnitude. In a.), the half-light radius
   $R_{\rm e}$ is plotted, in b.) and c.) the $V-I$ color and $B-V$
   color is plotted, d.)  shows the iron content [Fe/H]. In panel e.)
   and f.), the S\'ersic index $n$ and central surface brightness
   $\mu_{0}$ are plotted and in g.) the Faber-Jackson (FJ) relation is
   plotted. All these quantities are plotted against the $V$-band
   magnitude, except the FJ relation which are plotted as a function
   of the $B$-band luminosity. The isolated models are plotted by
   connected black diamonds with increasing final halo mass. For the
   merger trees the models are represented by different symbols
   depending on the final dark matter mass as indicated by the
   legend. Our models are compared with observational data obtained
   from \protect\citet{derijcke05} (DR05, dEs),
   \protect\citet{graham03} (GG03, dEs), \protect\citet{vanzee04}
   (dEs), \protect\citet{vanzee00} (dIrr), \protect\citet{hunter06}
   (dIrr), LG data (dSphs) come from \protect\citet{peletier93},
 \protect\citet{irwin95}, \protect\citet{saviane96},
 \protect\citet{grebel03}, \protect\citet{mcconnachie06},
 \protect\citet{mcconnachie07}, \protect\citet{zucker07}, Perseus data
 from \protect\citet{derijcke09} (dSphs/dEs), Antlia data from
 \protect\citet{castelli08} (dEs).  For the [Fe/H]$-M_{V}$ plot, data
 from \protect\citet{kirby13} was used for MW dSphs, M31 dSphs and
 Local Group dIrr. For the Faber-Jackson relation data from
 \protect\citet{geha03} (dEs), \protect\citet{kleyna05} (dSphs),
 \protect\citet{mateo98} (dEs/dSphs), \protect\citet{peterson93}
 (dEs), \protect\citet{vanzee04} (dEs) and \protect\citet{derijcke05}
 (dEs) is used. The dIrr, dSph and dE are shown by dots in different
 colors, respectively black (magenta), light-grey (yellow) and
 dark-grey (green) . (A color version of this figure is available in
 the online journal). \label{fig:bigFig}}
 \end{center}
 \end{minipage}
\end{figure*}

In panel a.) of Fig. \ref{fig:bigFig} the effective radius, $R_{e}$ is
shown as a function of the V-band magnitude. The black hexagons
represent the isolated simulations, where the increase in V-band
magnitude follows the increase in final halo mass (from respectively
10$^9$ M$_{\odot}$ to 10$^{10}$ M$_{\odot}$). The merger simulations
are indicated by different symbols/colors corresponding to different
final halo masses. In Table \ref{table_finalprop}, the value of the
effective radius is given for each of the simulated models.

In the observational data, the effective radius increases with
increasing V-band luminosity. The simulations have the same trend, but
the slope of the merger simulations is steeper compared with the
observational data with which the slope of the isolated models shows a
better agreement. However, our isolated simulations are too compact
while the merged galaxies are larger and more in agreement with
observed dwarf galaxies.

The larger effective radius of dwarf galaxies with merger histories
indicates that star formation is more widespread in these
galaxies. One important factor in determining the size of a galaxy's
stellar body is the depth of its DM potential, as this influences the
gravitational force on the gas. In the isolated models we already
notice a conversion of the cusped NFW profile to a cored density
profile due to baryonic processes \citep{cloetosselaer12}. In
paragraph \ref{subsection:DMhalo}, we show that the flattening of the
cusp is more pronounced in galaxies with a merger history. This may
explain the difference in $R_e$ between simulated galaxies with and
without merger histories in panel a.) of
Fig. \ref{fig:bigFig}. Moreover, the difference between the isolated
models and the merger models increases with halo mass. This may be due
to the fact that the dark matter density distribution flattens more in
more massive merger models.

\begin{table*}
 \begin{minipage}[ht]{\linewidth}
  \begin{center}
   \caption{Final properties of our simulations. The different blocks
     represent different final halo. Columns: (1) model type, (2) the
     final stellar mass in units of 10$^6$ M$_{\odot}$, (3) V-I color,
     (4) mass-weighted metallicity, (5) the S\'ersic index. (6) the
     mean surface brightness within the half-light radius, (7) the
     central one dimensional velocity dispersion, (8) the logarithm of
     the final specific stellar angular momentum, (9) the ellipticity,
     (10) the ratio of the maximal stellar velocity divided by the
     central one dimensional velocity of the galaxy and of a isotropic
     rotator, (11) and (12) respectively the amount of major and minor
     mergers in the formation histories, where we consider minor
     mergers to have mass ratios greater than
     1:3.}\label{table_finalprop} \begin{tabular}{lrrrrrrrrrrrrr}
\hline 
$\mathrm{model}$ & $M_{\rm \star,f}$ & $R_{\rm e}$ & $V-I$ & $[Fe/H]$ & $n$ & $\mu_{0,V}$& $\sigma_{\rm 1D,c}$ & $\log{j_{\star,f}}$ & $e$ & $(V/\sigma)^{\star}$ & $\#~MM$ & $\#~mM$  \\
 & $[10^{6}\ M_{\odot}]$ & $[kpc]$ & & & & $[mag]$ & $[km/s]$ & $[km~sec^{-1}~kpc]$ & & & & \\
\hline 
\hline 
\multicolumn{14}{l}{$\mathrm{M_{ h}=1 \times 10^{9}\ M_{\odot}}$} \\
\hline 
Isol & 5.61 & 0.21 & 0.83 & -1.46 & 1.00 & 22.81 & 8.54 & -0.29 & 0.26 & 0.57 & 0 & 0 \\
MT 1 & 1.36 & 0.24 & 0.71 & -1.23 & 0.98 & 23.71 & 10.01 & 0.25 & 0.42 & 1.12 & 4 & 5 \\
MT 2 & 2.71 & 0.21 & 0.82 & -1.28 & 0.80 & 24.05 & 8.18 & -0.74 & 0.11 & 0.36 & 3 & 8 \\
MT 3 & 1.25 & 0.20 & 0.81 & -1.24 & 0.78 & 24.51 & 7.85 & -0.35 & 0.32 & 0.59 & 6 & 9 \\
MT 4 & 0.79 & 0.20 & 0.80 & -1.16 & 0.94 & 24.46 & 7.76 & -0.39 & 0.34 & 0.53 & 4 & 12 \\
\hline 
\multicolumn{14}{l}{$\mathrm{M_{ h}=2.5\times 10^{9}\ M_{\odot}}$} \\
\hline 
Isol & 27.22 & 0.33 & 0.84 & -1.80 & 1.10 & 21.94 & 11.99 & 0.57 & 0.12 & 2.05 & 0 & 0 \\
MT 1 & 16.97 & 0.52 & 0.83 & -1.33 & 0.66 & 23.99 & 9.27 & 0.64 & 0.20 & 1.46 & 4 & 8 \\
MT 2 & 33.13 & 0.84 & 0.81 & -1.10 & 0.32 & 24.49 & 13.39 & 1.03 & 0.22 & 1.89 & 3 & 7 \\
MT 3 & 9.22 & 0.29 & 0.81 & -1.66 & 0.99 & 23.04 & 11.00 & 0.24 & 0.16 & 0.56 & 11 & 8 \\
MT 4 & 13.40 & 0.64 & 0.79 & -1.38 & 0.88 & 24.46 & 14.14 & 0.39 & 0.43 & 0.26 & 4 & 13 \\
MT 5 & 12.97 & 0.52 & 0.75 & -1.31 & 0.60 & 23.01 & 14.39 & 0.42 & 0.11 & 0.52 & 6 & 8 \\
\hline 
\multicolumn{14}{l}{$\mathrm{M_{ h}=5 \times 10^{9}\ M_{\odot}}$} \\
\hline 
Isol & 94.86 & 0.43 & 0.84 & -1.56 & 0.77 & 21.82 & 19.69 & 0.69 & 0.22 & 1.14 & 0 & 0 \\
MT 1 & 55.50 & 0.90 & 0.82 & -1.30 & 0.61 & 24.03 & 12.35 & 0.49 & 0.06 & 0.50 & 7 & 12 \\
MT 2 & 65.50 & 1.13 & 0.83 & -1.25 & 0.52 & 24.45 & 11.00 & 0.11 & 0.31 & 0.19 & 5 & 17 \\
\hline 
\multicolumn{14}{l}{$\mathrm{M_{ h}=7.5 \times 10^{9}\ M_{\odot}}$} \\
\hline 
Isol & 233.56 & 0.49 & 0.86 & -1.21 & 0.90 & 20.93 & 25.33 & 0.71 & 0.10 & 1.23 & 0 & 0 \\
MT 1 & 68.70 & 1.15 & 0.79 & -1.50 & 0.18 & 24.75 & 15.73 & 1.02 & 0.26 & 0.40 & 12 & 8 \\
MT 2 & 127.58 & 1.78 & 0.82 & -1.10 & 0.45 & 24.90 & 18.38 & 1.42 & 0.21 & 0.86 & 5 & 15 \\
\hline 
\multicolumn{14}{l}{$\mathrm{M_{ h}=1 \times 10^{10}\ M_{\odot}}$} \\
\hline 
Isol & 1288.10 & 0.41 & 0.95 & -0.39 & 0.89 & 18.79 & 58.69 & 0.40 & 0.16 & 0.17 & 0 & 0 \\
MT 1 & 178.56 & 1.68 & 0.85 & -1.17 & 0.61 & 24.12 & 18.42 & 1.31 & 0.47 & 0.45 & 5 & 7 \\
\hline 
\end{tabular}

  \end{center}
 \end{minipage}
\end{table*}

\subsubsection{The V$-$I and B$-$V color.}
The V$-$I and B$-$V color as a function of the V-band magnitude is
shown in respectively panel b) and c) of Fig. \ref{fig:bigFig}.  The
V$-$I color has a value significantly below V$-$I$\sim 0.7$~mag only
in stellar populations with ages below a few 100 Myr. We see that the
V$-$I color is constant around a value of $\sim 0.8$~mag for the
entire magnitude range with a scatter of 0.05~mag. There are some
galaxies which, due to a late star formation burst, have smaller
values for V$-$I. For example, MT5 from Fig. \ref{fig:SFR} has the
lowest V$-$I value in its mass range (see Table
\ref{table_finalprop}), due to a recent peak in star formation at 12.4
Gyr. The other MTs have similar values for their V$-$I color which is
due either to a similar late SF peak or by continuous star formation
at a small rate. Likewise, the B$-$V color scatters within 0.1~mag
around a value of $\sim 0.5$~mag for the entire luminosity range.

In terms of the V$-$I color, the simulations are significantly bluer
than dSphs and dEs, with the merger simulations slightly bluer than
the isolated simulations. In terms of the B$-$V color, the simulations
are comparable to observed dIrrs, although on the red side of the dIrr
color distribution.

This is caused by the larger gas fraction of the simulated dwarfs
since there are no environmental effects present that could remove
gas. As a result, low level star formation occurs during the last 6
Gyr of the simulations generating young, blue stars. The isolated
galaxies are generally redder as most of their stars have been formed
early on in the simulation. Overall, very different SFHs due to
different merger histories all result in approximately the same V$-$I
and B$-$V color.

Our rather blue models, which are more in agreement with observed dIrr
due to their large gas content and ongoing star formation, could be
transformed into red and dead dSph by external interventions which
remove the gas and shut down star formation for the last Gyr of the
simulation. Examples of such external processes are: ram-pressure
stripping \citep{mayer06, boselli08}, tidal interactions
\citep{mayer01, mayer01b}, and the UV background
\citep{shaviv03}. However, for the simulations presented here we did
not implement such external processes, so star formation continues
till $z=0$ and the simulated galaxies are bluer than the observed
dSph/dE galaxies. and more in agreement with the observed dIrrs which
are caracterized by ongoing star formation.

Incorporating gas depleting processes, such as gas stripping and the
cosmic UV background, in the simulation code is currently ongoing,
using the cooling and heating curves from \cite{derijcke13} which
contain the UVB.

\subsubsection{The metallicity}
Panel d) in Fig \ref{fig:bigFig} shows the luminosity weighted Iron
abundance, [Fe/H], a tracer of the metallicity of the stars, as a
function of the V-band magnitude. The simulations are compared with
observational data from \cite{kirby13} for Local Group dIrr
(black/magenta dots), M31 dSph (light-grey stars/orange dots), and
Milky Way dSph (ligth-grey/yellow dots).

With increasing mass, the isolated simulations tend to become too
compact leading to fast and self-enriching star formation. This causes
the bright side of their M$_V$-[Fe/H] relation to be too steep. For
the least massive merger models, with halo masses around 10$^{9}$
M$_{\odot}$, the metallicity is too high by about 0.4 dex compared
with the observational data. This is because only near the galaxy
center does the gas density exceed the density threshold for star
formation. This centrally concentrated star formation then
self-enriches too much. Star formation is centrally concentrated in
the least massive isolated model as well but in this case a
significant fraction of the stars form early on from almost unenriched
gas, causing the mean metallicity to be lower than in the merger
models.

The metallicities of the more massive merger models are in better
agreement with the observations. In the latter, star formation occurs
spatially more widespread and self-enriches less. Except for the least
massive ones, the metallicities of the merger simulations compare well
with those of the observed Local Group dwarfs.

\begin{figure}
 \centering
 \includegraphics[width=0.5\textwidth]{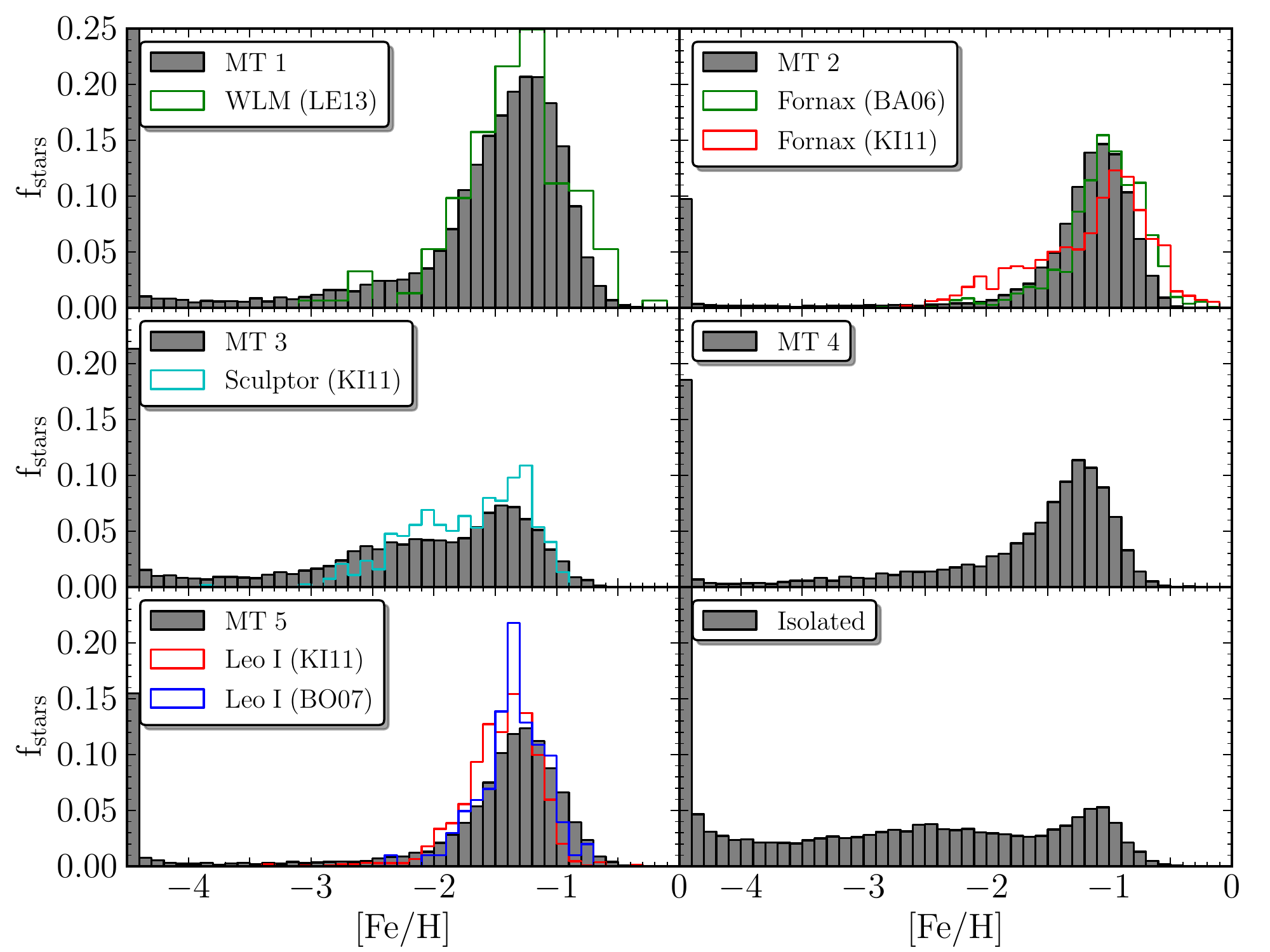}
 \caption{A histogram of the metallicity distribution of the
   stars. The simulations all produce dwarf galaxies with final halo
   mass of 2.5$\times$10$^9$ M$_{\odot}$. The simulations are compared
   with observational data from \protect\cite{battaglia06} (BA06),
   \protect\cite{kirby11_3} (KI11) and \protect\cite{bosler07}
   (BO07). (A color version of this figure is available in the online
   journal.) \label{fig:metall}}
\end{figure}

In Fig. \ref{fig:metall}, the metallicity distribution function (MDF)
of the different merger trees and the isolated simulations at $z=0$ is
plotted in a histogram. As explained in paragraph \ref{subsection:IC},
the gas already has a very small metallicity ($Z=10^{-4}\ Z_{\odot}$)
from the start of the simulation. All stars formed from this gas will
have exactly the same Iron abundance of [Fe/H]$=-4.45$, causing a
spike in the MDF at this metallicity. This is an artefact of our
idealised initial conditions.

The isolated simulated galaxies generally have a larger fraction of
metal-poor stars than galaxies with a merger history. This is due to
the first, large star formation peak consuming the metal-poor gas
reservoir in isolated simulated galaxies. In merger simulations, the
interaction induced star formation rapidly boosts the metallicity to
[Fe/H]$\sim -1$, thus suppressing the low-metallicity tail. One
notable exception is MT3 whose merger tree contains relatively few
mergers during the first 3~Gyr. Star formation in its isolated
progenitor galaxies produces a population of low-metallicity stars,
peaking in the metallicity range [Fe/H]$\sim -3$ to $-2$.

As an illustration, we compare the MDFs of some of the merger tree
simulations with those of observed Local Group dwarf galaxies:~Fornax,
Sculptor, Leo~I and WLM. With absolute magnitudes of, respectively,
M$_V=-13.3$~mag, $-11.1$~mag, $-11.8$~mag and $-14.92$~mag
\citep{irwin95,mateo98,lokas09,devaucouleurs91} these galaxies fall in
the luminosity interval covered by the simulations. Each galaxy is
compared with a merger tree simulation that closely matches its mean
[Fe/H]:~[Fe/H]=-0.99, -1.68, and -1.43 for Fornax, Sculptor, and Leo~I
\citep{kirby11_3} and WLM with ~[Fe/H]=-1.28 \citep{leaman13},
respectively, and [Fe/H]=-1.1, -1.66, -1.31, and -1.33 for MT2, MT3,
MT5, and MT1 respectively.  All MDFs, simulated and observed, are
normalized to unity over the metallicity interval from [Fe/H]=-4 to
[Fe/H]=0. The Fornax MDFs are taken from \cite{battaglia06} (BA06) and
\cite{kirby11_3} (KI11), those of Sculptor from KI11, those of Leo~I
from KI11 and from \cite{bosler07} (BO07) and the WLM MDF is taken
from \cite{leaman13}. Clearly, there can be a significant
author-to-author difference between observed MDFs of the same
galaxy. This could be caused by the different number of stars included
in the samples and by the different spatial extents covered by the
observations, if the stellar populations are spatially segregated.

The low-metallicity tail of the Fornax data of BA06 was explained to
originate from the infall of a metal poor component based on its
non-equilibrium kinematics and radial metallicity gradient. The
analysis of KI11 suggested Fornax to have an extended SFH due to the
metal-rich peak in the MDF which is caused by star formation from
previously enriched gas. As MT2 already has a massive component
containing roughly half of its final mass at one Gyr in the simulation
after which it receives metal-poor gas from multiple minor mergers
triggering SF, this explains why the simulation is most in line with
the results of KI11. The same is true for the MDFs of MT1 and MT4.

Like MT3, Sculptor has a bimodal MDF with a high-metallicity peak
around [Fe/H]$\sim -1.3$ and a low-metallicity peak below [Fe/H]$\sim
-2$. We merely wish to illustrate that such bimodal MDFs are also
found among real dwarfs in this luminosity range. Moreover, the
agreement is not perfect: MT3 contains more low-metallicity stars than
Sculptor and the high-metallicity peak is less strong.

We can conclude from Fig. \ref{fig:metall} that the peaked MDFs of the
merger-tree simulations are much more in agreement with the
observations than the flat MDFs of the isolated models.

\subsubsection{The S\'ersic parameters}
Panels e) and f) show in Fig. \ref{fig:bigFig} show respectively the
S\'ersic index $n$ and the central surface brightness in the V-band,
$\mu_{0,V}$. In both cases the simulations are compared to
observational data of dE/dSph galaxies. Generally, for all mass models
and different merger trees, the models overlap with the S\'ersic
  parameter data in the regime of the dwarf spheroidals. The scatter
in the simulations is mainly due to the different merger histories.

\begin{description}
\item[\textbf{S\'ersic index, $n$}] The merger simulations have
  similar $n$-values as the observational data and are smaller or
  equal compared to the S\'ersic indices of isolated simulations. We
  see that neither the merger tree, nor the flattening of the central
  core has a significant influence on the S\'ersic index of the
  simulations.
\item[\textbf{Central surface brightness in the V-band, $\mu_{0,V}$}]
  The isolated models follow the trend of increasing central surface
  brightness with increasing V-band magnitude. The most massive merger
  simulations have lower central surface brightnesses compared to the
  isolated models but are still in agreement with the
  observations. This low $\mu_{0,V}$-value is due to the flatter DM
  core which causes star formation to occur less centrally and results
  in a more extended, less dense stellar body.
\end{description}

\subsubsection{The Faber-Jackson relation}\label{subsection:sigma}
After the photometric relations, we now turn to the kinematic
relations, where panel g.) of Fig. \ref{fig:bigFig} shows the velocity
dispersion of the stars as a function of the B-band luminosity. Our
simulations follow the same trend as the observational data. The
isolated simulations have velocity dispersions which are somewhat
large compared to the observations. The merger simulations are better
in agreement with the observations but they show some spread which is
due to the different merger histories which result in different
flattenings of the dark matter core.

For more massive haloes, the velocity dispersion of the merged
galaxies deviates markedly from that of the isolated galaxies. Again,
this is most likely due to the stronger flattening of the dark matter
halo in more massive galaxies, see section
\ref{subsection:DMhalo}. Due to the lower central density and the
shallower potential, stellar velocities will be lower, resulting in a
lower velocity dispersion.

\section{Discussion}
\label{section:results}

\subsection{The dark matter halo}
\label{subsection:DMhalo}

\begin{figure}
 \centering
 \includegraphics[width=\columnwidth]{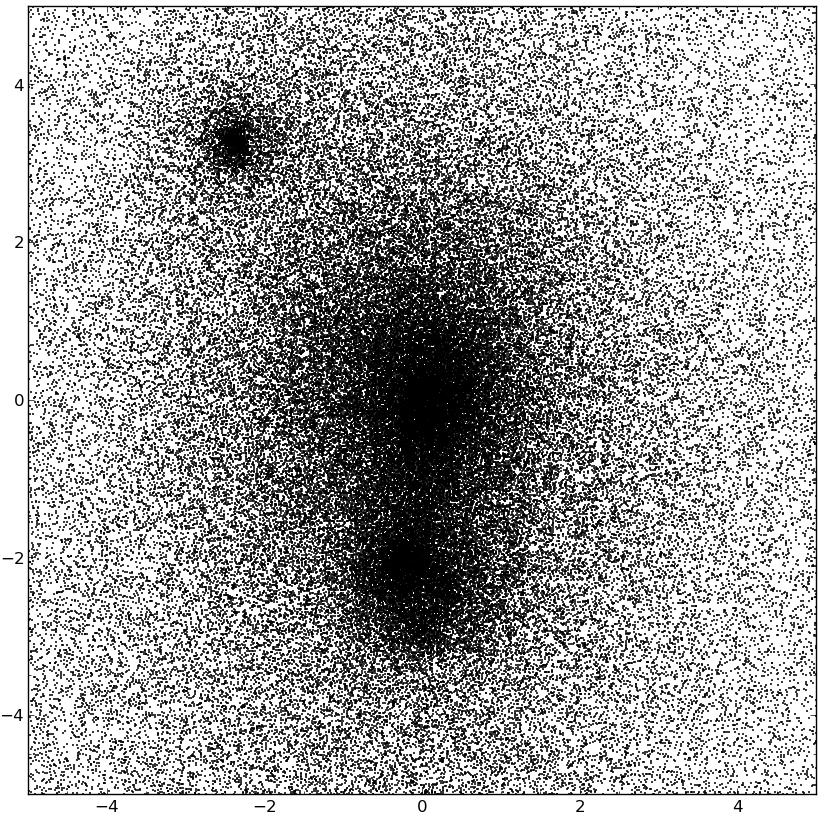}
 \caption{The dark matter distribution of a halo with mass
   2.5$\times$10$^{9}$ M$_{\odot}$ at $z$=0. The units on the the box
   are in kpc.\label{fig:DMsnapshot}}
\end{figure}

\begin{figure}
 \centering
 \includegraphics[width=\columnwidth]{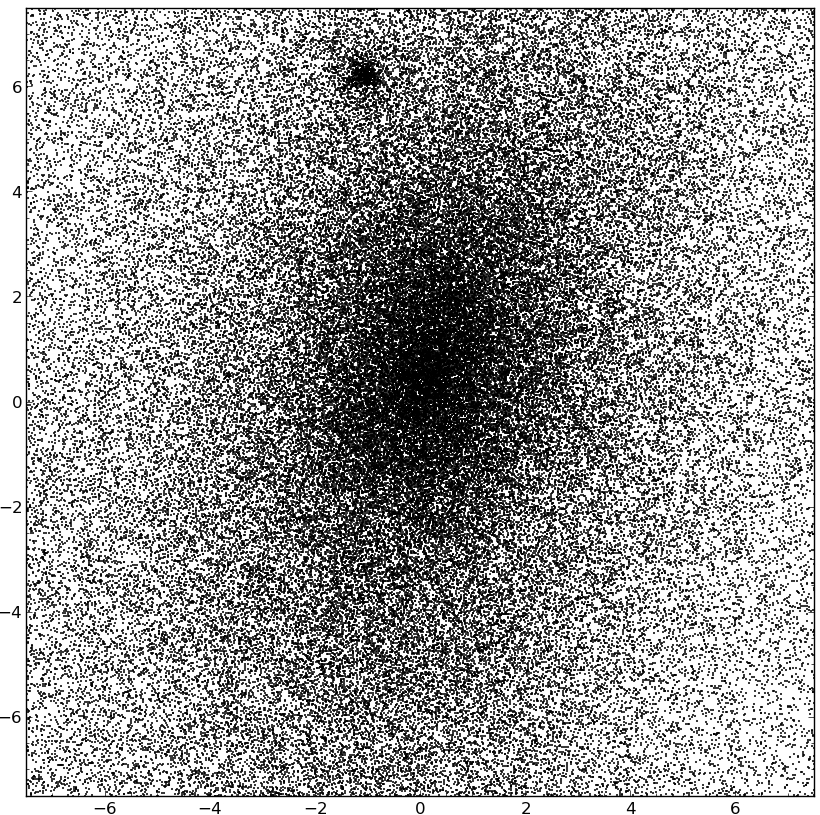}
 \caption{The dark matter distribution of a halo with mass
   7.5$\times$10$^{9}$ M$_{\odot}$ at $z$=0. The units of the box are
   in kpc. \label{fig:DMsnapshot2}}
\end{figure}

In order to create dark-matter density profiles, the isolated
simulations are centered on the center of mass of the dark matter
haloes. For the merger simulations, this approach is less obvious
since the merging process of the dark matter haloes produces tidal
tails which extend to large radii. As a result, the center of mass of
a halo can deviate significantly from what one would consider ``by
eye'' to be the location of the center of the main halo. Moreover, as
can be seen in Fig. \ref{fig:DMsnapshot} and
Fig. \ref{fig:DMsnapshot2}, the dark-matter haloes of galaxies with a
merger history contain substantial substructure until the end of the
simulation. So, to find the center of the main halo, a 3D Voronoi
tessellation \citep{00schaap} is used to determine the density at the
position of each dark matter particle. The 15 particles with the
highest densities are then selected. Finally, the center of the halo
is equated to that particle out of those 15 which has the largest mass
within a 2~kpc sphere in order to avoid local density peaks. A visual
check proved that this procedure yields a meaningful estimate for the
halo center. The dark-matter density profile is derived from the mass
enclosed inside increasingly large spherical shells centered on the
halo center identified as explained above.

\begin{figure}
 \centering
 \includegraphics[width=\columnwidth]{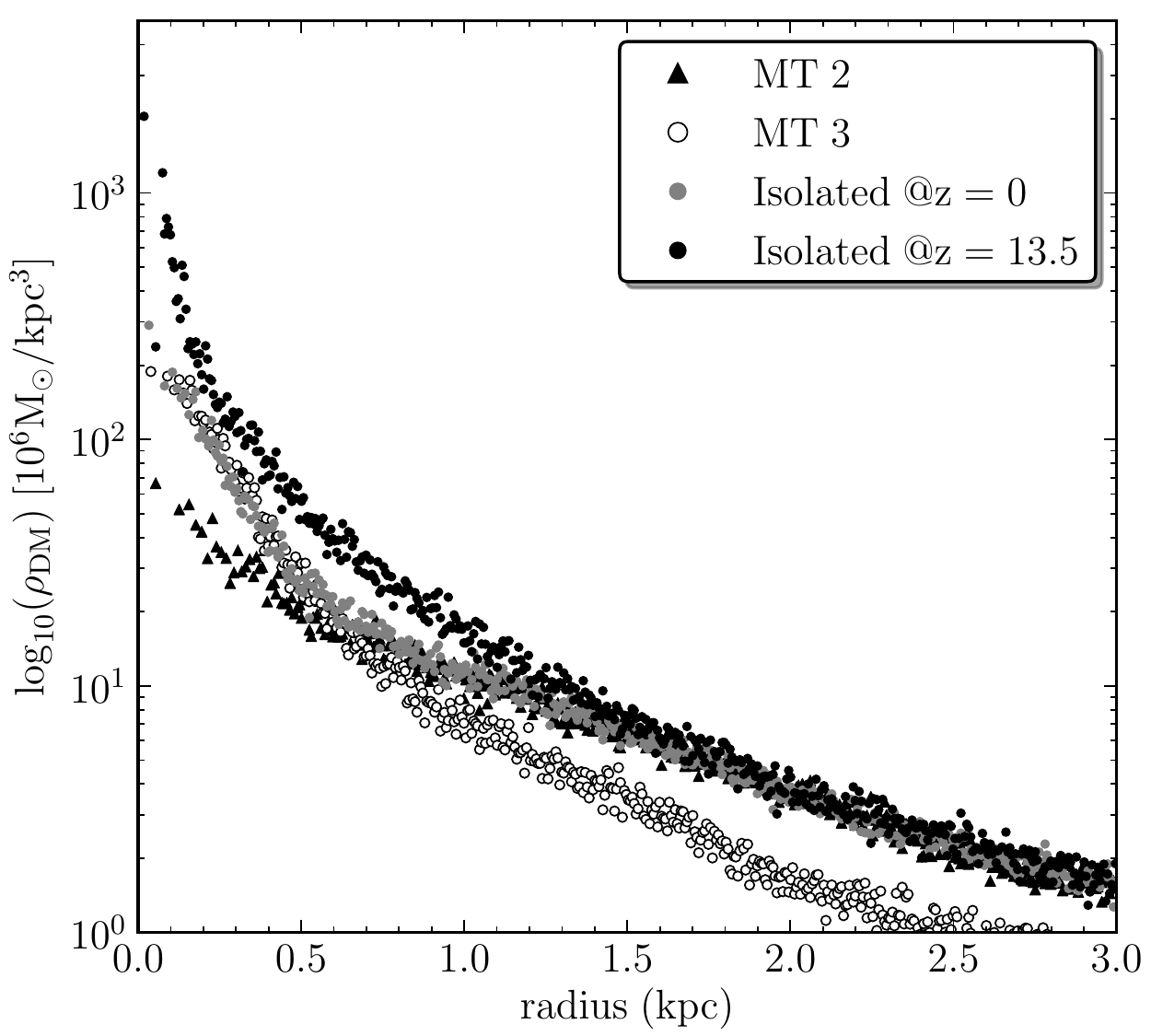}
 \caption{The dark matter density profile of the merger simulations
   MT2 and MT3 at $z=0$ and one isolated galaxy with a final
   halo mass of 2.5$\times$10$^{9}$ M$_{\odot}$ at $z=0$ and
   $z=13.5$. \label{fig:DMsmall}}
\end{figure}

\begin{figure}
 \centering
 \includegraphics[width=\columnwidth]{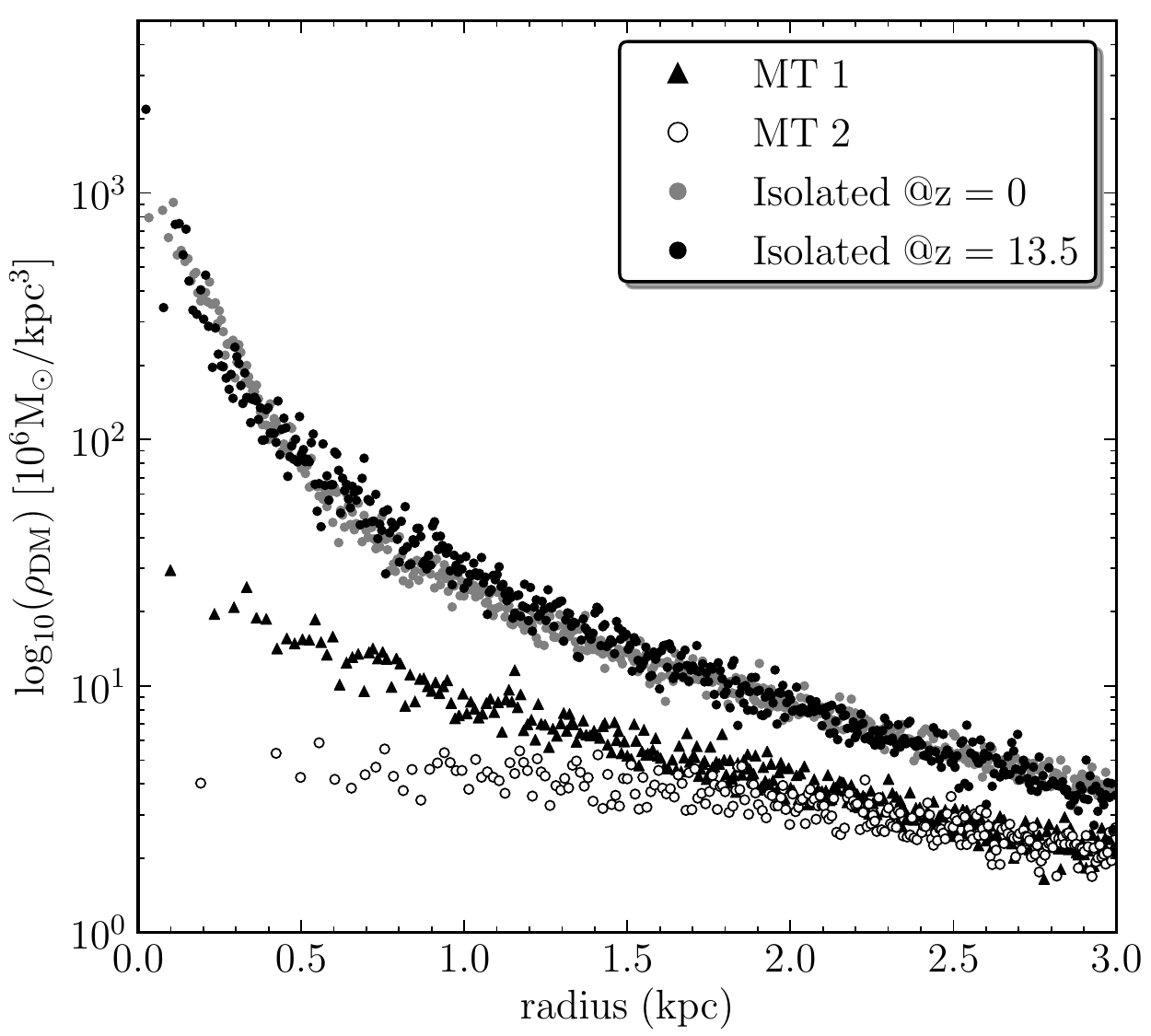}
 \caption{The dark matter density profile of the merger simulations at
   $z=0$ and one isolated galaxy with a final halo mass of
   7.5$\times$10$^{9}$ M$_{\odot}$ at $z=0$ and
   $z=13.5$. \label{fig:DMbig}}
\end{figure}

In Figs. \ref{fig:DMsmall} and \ref{fig:DMbig}, the density profiles
of the merged haloes with halo mass of respectively
2.5$\times$10$^{9}$ M$_{\odot}$ and 7.5$\times$10$^{9}$ M$_{\odot}$
are plotted at $z=0$ together with the density profiles of an equally
massive isolated galaxy at $z=0$ and $z=13.5$. Very often, we see a
conversion from a cusped NFW-profile (black dots) to a cored, or at
least less steep, density profile (light-grey (cyan) dots) in the
isolated galaxy simulations due to the effects of stellar feedback
\citep{cloetosselaer12}. Haloes of galaxies with a merger history
appear to become even shallower by $z=0$ than the isolated ones.
Comparing the dark-matter density profiles in Figs. \ref{fig:DMsmall}
and \ref{fig:DMbig} with the merger trees that produced them (see
e.g. Fig. \ref{fig:mergers_all}) and the corresponding star-formation
histories (see e.g. Figs \ref{fig:SFR} and \ref{fig:SFR2}) shows that
the former is a non-trivial function of the merger history and the
baryonic processes. The absorption of the orbital energy involved in a
major merger will tend to inflate the dark-matter halo, causing the
cusp to weaken. If a merger causes a rapid inflow of gas, it may
compress the cusp whereas gas expulsion by supernovae may weaken the
cusp.

\begin{figure}
 \centering
 \includegraphics[width=\columnwidth]{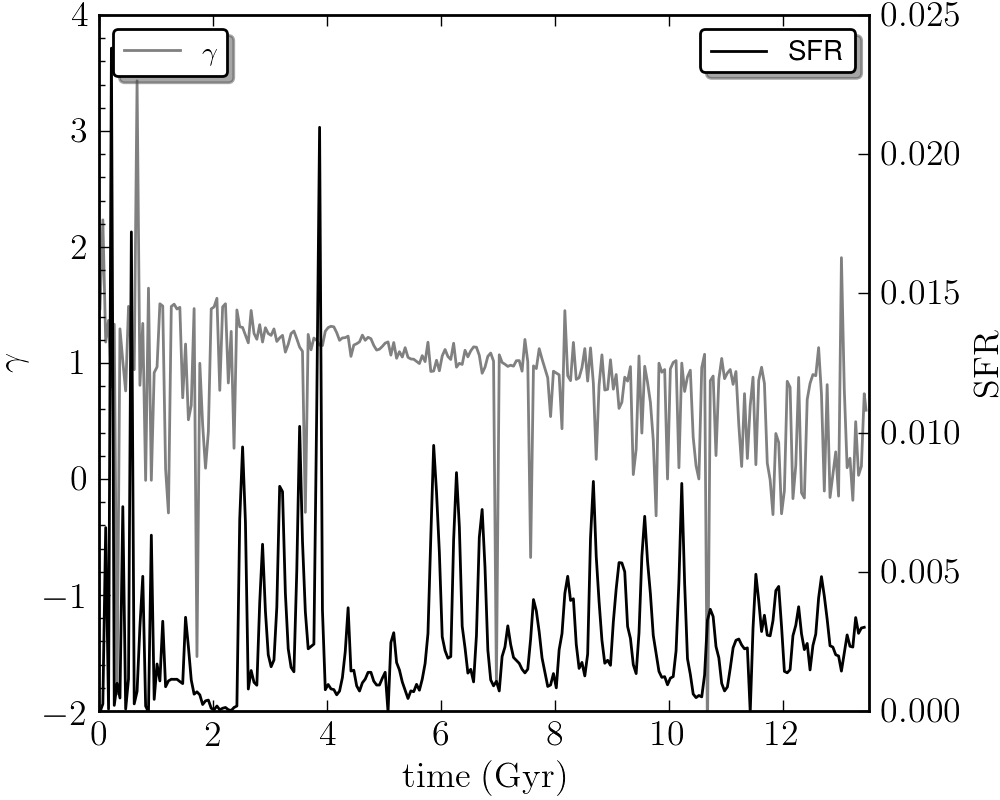}
 \caption{The evolution of the slope of the most massive component of
   MT2 in grey. The SFR is plotted by the black line. \label{fig:gammaMT2}}
\end{figure}

\begin{figure}
 \centering
 \includegraphics[width=\columnwidth]{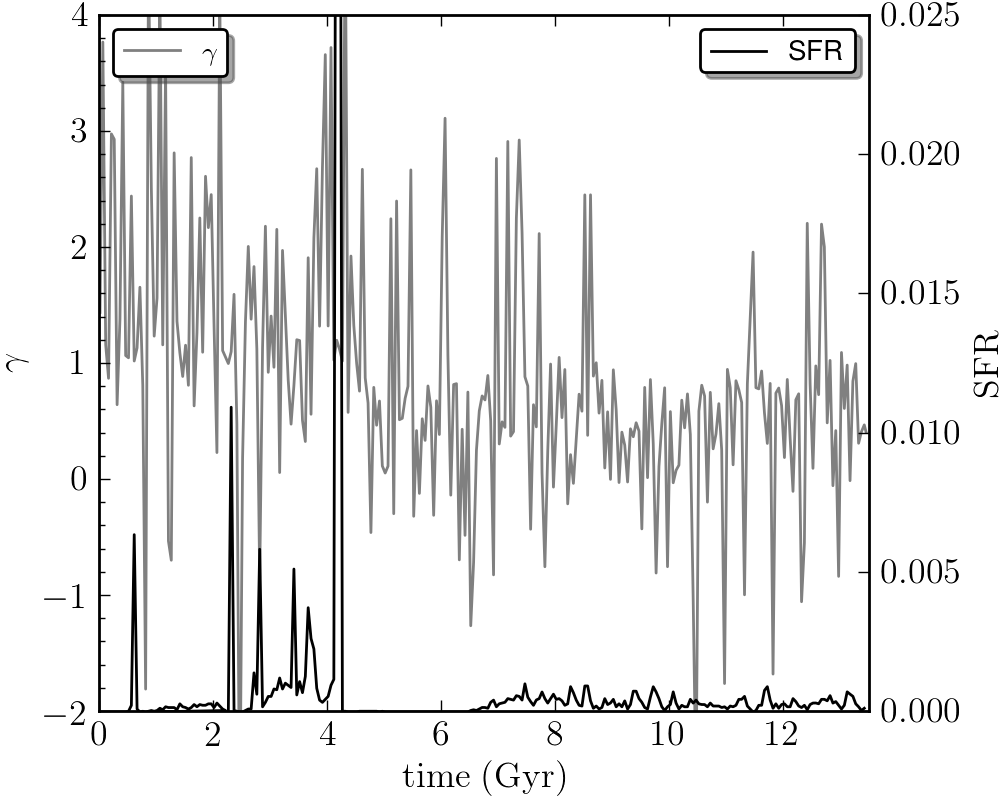}
 \caption{The evolution of the slope of the most massive component of
   MT3 in grey. The SFR is plotted by the black line.\label{fig:gammaMT3}}
\end{figure}

In Fig. \ref{fig:gammaMT2} and Fig. \ref{fig:gammaMT3} the evolution
of the inner slope of the dark-matter density profile, denoted by
$\gamma$, is plotted for respectively MT2 and MT3 of the merger models
with final halo mass of
M$_{h,f}$=2.5$\times$10$^{9}$M$_{\odot}$. $\gamma$ is determined by a
least-square fit of a Nuker law \citep{lauer95} to the density profile:
\begin{equation}
 \rho(r)= \frac{\rho_s}{r^\gamma(r^\beta+r_s^\beta)^{\alpha}}.
\end{equation}
This function corresponds with a broken power-law function, where the
break radius is described by $r_s$ and the sharpness of the transition
is set by $\beta$. $\gamma$ represents the slope in the inner part,
e.g. for $r \ll r_s$, $\rho(r) \sim r^{-\gamma}$, and $\alpha$
determines the outer power law as for $r \gg r_s$ $\rho(r) \sim
r^{-\alpha\beta-\gamma}$. The NFW profile corresponds to $\gamma=1$, $\beta=1$, and
$\alpha=2$. We omit the inner 60~pc from the fit; this corresponds to
twice the gravitational softening length.

MT2 is a merger tree which already starts with a quite massive halo at
early times, e.g. $\sim$53\% of the final halo mass is present in the
main halo after one Gyr in the simulation and the halo grows by the
subsequent addition of minor mergers, each containing $\sim$1\%-10\%
of the final halo mass. In Fig. \ref{fig:gammaMT2}, one first notices
the adiabatic contraction of the halo, signaled by an increase of
$\gamma$ to values above 1, due to the initial collapse of gas in the
dark-matter potential well. Afterwards, the slope gradually decreases
again due to the rapid removal of gas from the inner regions during
repeated small starbursts triggered by the many minor merger
events. MT1 and MT4 show a similar behaviour as MT2 in this regard.


MT3, on the other hand, is a merger tree where the mass builds up
slowly over time, e.g. only $\sim$33\% of the final halo mass is
present in the main halo after one Gyr. Here, we trace the evolution
of the slope $\gamma$ in the most massive halo present at each point
of time in the tree MT3. A major merger occurs at 4 Gyr resulting in a
large star formation peak and a subsequent shutdown of star
formation. As shown in Fig. \ref{fig:gammaMT3}, the slope initially
rapidly increases above 1, as in MT2. In this case, however, star
formation is very low powered and the dark-matter cusp appears quite
resilient against any small-scale gas motions. Only after the steep
increase of the star-formation rate and the feedback activity
connected with the major merger around 4~Gyr does the slope drop below
$\gamma=1$. The starburst is actually so strong that star formation is
halted for the next 1.5 Gyr and, when restarted, remains very weak and
unable to further affect the dark-matter profile. In this case, the
inner dark-matter slope remains stable. The dark-matter density
profile of MT5 behaves similarly to that of MT3.

As shown in the literature, baryonic processes can explain the
discrepancy between the cored dark matter density profiles of observed
galaxies and the cusped dark matter density profiles deduced from
cosmological simulations. First, the rapid removal of gas due to
stellar feedback results in a non-adiabatically response of the dark
matter halo and introduces a flattening of the cusped dark matter halo
\citep{navarro96b, read05, mashchenko06, governato10, pontzen12,
  governato12, cloetosselaer12, brooks14}. Secondly, the transfer of
energy and/or angular momentum to the dark matter by infalling objects
can transfer the cusped inner dark matter density profile into a more
cored profile \citep{goerdt06, goerdt10, cole11}. Repeated minor
mergers can trigger small starbursts that rapidly evacuate gas from
the galaxy center, with each burst slightly lowering the slope of the
dark-matter density profile. Strong starbursts caused by major mergers
have the same effect but, during star-formation lulls, the dark-matter
density profile remains stable.

Merger trees with more massive final haloes show the same behavior,
although Fig. \ref{fig:DMbig}, which shows the density profiles of
haloes with final mass $M=7.5\times 10^9$~M$_\odot$, suggests that the
flattening effect is much more pronounced for higher masses. This is
likely caused by the stronger fluctuations in the star-formation rate
(see Fig. \ref{fig:SFR2}) and by the fact that more massive merging
galaxies need to absorb more orbital kinetic energy. Tree MT2 in this
mass series of simulations contains a massive progenitor already early
on which grows mainly through minor mergers. Star formation continues
throughout the simulation, constantly reducing the inner dark-matter
slope. MT1 contains major mergers that stop star formation for
considerable timespans, limiting the flattening of the dark-matter
cusp.

We conclude that the dark matter haloes are strongly influenced by the
merger history and the resulting baryonic processes.  \cite{oh11a}
determined the inner density slopes within the central kiloparsec of
the THINGS dwarf galaxies. They define the inner density slope
$\alpha$ from a fit to the density profile $\rho \propto 1/r^{\alpha}$
and concluded a value of $\alpha=0.29 \pm 0.07$ for the THINGS dwarf
galaxies sample. In \cite{oh11b}, the inner density profiles of
simulated dwarf galaxies in a cosmological simulation were reported to
have a slope of $\alpha=0.4 \pm 0.1$.  Here, we find, for instance,
that for the inner dark-matter density slope of the haloes with final
mass $M=2.5\times 10^9$~M$_\odot$ $\alpha=0.56 \pm 0.27$.  In order to
explain the different slopes of the equally massive dark matter
haloes, we have shown that small star formation peaks, due to repeated
minor mergers, are efficient at lowering the slope. Major mergers
cause a spike in the star-formation rate but feedback rapidly shuts
down star formation, thus limiting the effect on the inner dark-matter
density profile slope. This effect happens over all the full mass
range but is more pronounced in the more massive models. The resulting
shallow gravitational potential probably explains the large effective
radii (see subsection \ref{subsection:Re}) and low central velocity
dispersions (see subsection \ref{subsection:sigma}) observed in some
of the simulated galaxies.

\subsection{Stellar specific angular momentum}
\begin{figure}
 \centering \includegraphics[width=\columnwidth]{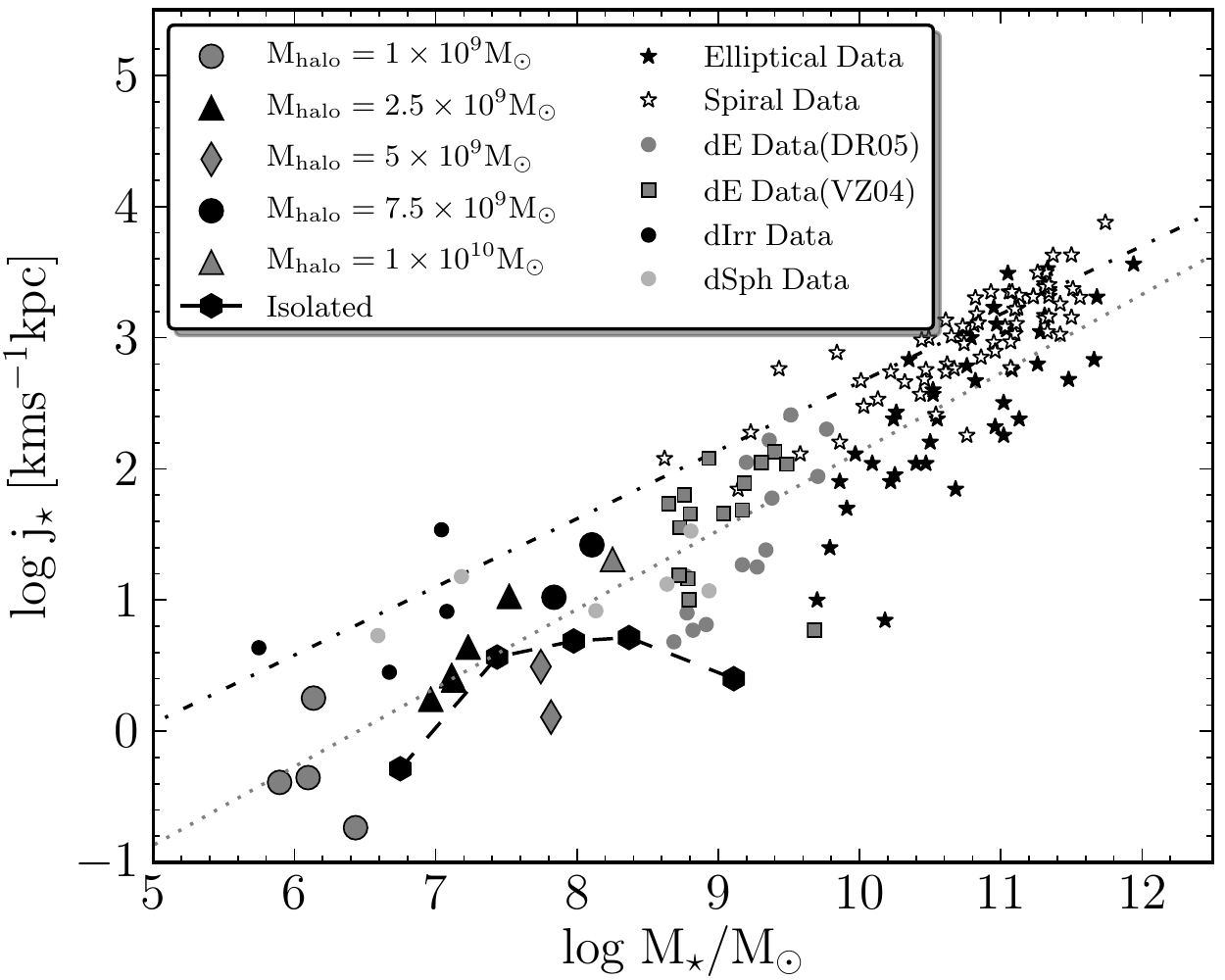}
 \caption{The stellar specific angular momentum at the end of the
   simulations ($z=0$) as a function of the stellar
   mass. Observational data of elliptical galaxies and spiral
     galaxies are taken from \protect\cite{romanowsky12} and a fit to
     the datapoints is plotted by a respectively the dotted and
     dash-dotted line. dE data of \protect\cite{derijcke05}(DR05) and
     \protect\cite{vanzee04}(VZ04) are plotted next to dIrr data and dSph
     data from \protect\cite{leaman12}, \protect\cite{kirby12},
     \protect\cite{kirby14}, \protect\cite{mcconnachie12},
     \protect\cite{derijcke06}, \protect\cite{mcconnachie06},
     \protect\cite{worthey04}, and
     \protect\cite{hidalgo13}.  \label{fig:angularMomentum}}
\end{figure}


\begin{figure}
 \centering
 \includegraphics[width=\columnwidth]{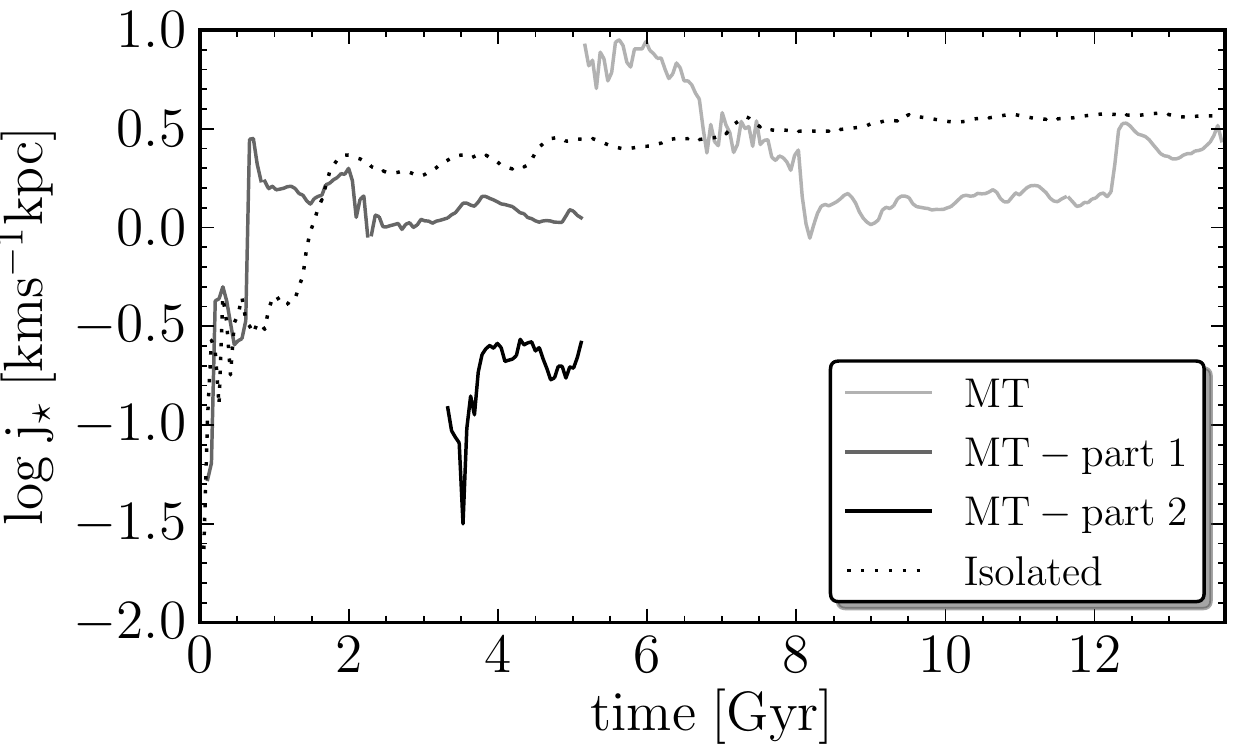}
 \caption{The evolution of the stellar specific angular momentum in
   time for an isolated simulation (dotted line) and for the merger
   simulation MT4, both with M$_{h,f}$=2.5$\times$10$^{9}$M$_{\odot}$
   (black, dark-grey and light-grey line). \label{fig:angularMomentumTime}}
\end{figure}

Fig. \ref{fig:angularMomentum} shows the specific angular momentum of
the stars, $j_{\star}$, calculated as the length of the vector sum of
the angular momenta of all the stars, using the center of mass of the
most massive stellar body as a reference point, divided by the total
stellar mass, as a function of the stellar mass, $M_{\star}$, at
$z=0$. The isolated simulations are represented by connected black
hexagons, with increasing stellar mass corresponding to increasing
halo mass. The merger simulations, shown as indicated by the legend,
represent different final masses of the dark matter halo. For
comparison, observational data are also plotted. The data of the
spiral and elliptical galaxies are taken from \cite{romanowsky12}, the
observational data of dE from \cite{derijcke05}(DR05) and
\cite{vanzee04}(VZ04) and data for dIrr and dSph are taken from
\cite{worthey04}, \cite{derijcke06}, \cite{mcconnachie06},
\cite{leaman12}, \cite{kirby12}, \cite{mcconnachie12},
\cite{hidalgo13}, and \cite{kirby14}. The simulations have similar
specific angular momentum as the observed dwarf galaxies.

Like the observed galaxies, the simulated galaxies follow a trend of
increasing stellar specific angular momentum with increasing stellar
mass. At a given halo mass, the scatter on $j_\star$, caused by the
different merger histories and star-formation histories, can be as
large as an order of magnitude. In particular, merger histories that
involve many mergers tend to produce galaxies with small $j_\star$
since the angular momenta of these mergers can cancel each
other. Merger trees that involve few mergers have less opportunities
for canceling orbital angular momenta and can produce galaxies with
high $j_\star$.

In Fig. \ref{fig:angularMomentumTime} the evolution of the stellar
specific angular momentum is shown for an isolated simulation
(dotted line) and a merger simulation (black,
  dark-grey and light-grey line), both with a final halo mass of
2.5$\times$10$^{9}$ M$_{\odot}$.
\begin{description}
\item[\textbf{Isolated galaxies}] The stochastic nature of star
  formation is responsible for most of the stellar angular momentum
  that is created during the first Gyr of the simulation as stars are
  not created in a perfectly spherically symmetric way. In
  Fig. \ref{fig:angularMomentumTime}, the black line shows the
  evolution of the stellar specific angular momentum of an isolated
  galaxy. During the first Gyr, a large increase in stellar specific
  angular momentum occurs due to a large star formation
  peak. In the next Gyr, stars are mainly born out of the
    turbulent ISM. As the stars inherit the kinematics of the gas
    particles they are born from, the specific stellar angular
    momentum will increase. In addition, SNIa feedback asymmetrically
    accelerates the gas and, as a reaction, affects the stellar
    motions, increasing the specific stellar angular momentum.

\item[\textbf{Merged galaxies}] The evolution of the stellar specific
  angular momentum for a merger simulation is plotted in Fig.
  \ref{fig:angularMomentumTime}. Around 5~Gyr into the simulation, two
  branches of the merger tree, represented by the black and dark-grey
  line, are put together. The joint system continues as indicated by
  the light-grey line. The large increase of $j_{\star}$ is the result
  of the vector sum of both the initial $j_{\star}$s of the branches,
  together with their orbital angular momentum. The incoming galaxy
  passes by the main galaxy at $\sim$6.4 Gyrs and starts to return to
  the main galaxy at $\sim$6.8 Gyr. At 7.8 Gyr it actually merges with
  the main galaxy, causing a large peak in the star formation rate
  (see Fig. \ref{fig:SFR}). As the SF is centrally concentrated, it
  will not change the angular momentum much but the stellar mass will
  increase, resulting in a net decrease of the specific angular
  momentum. The same happens during the first two SF peaks of MT4 (see
  Fig. \ref{fig:SFR}) at 2 and 2.2 Gyr which correspond to two
  decreases in $j_{\star}$ in the dark-grey curve in
  Fig. \ref{fig:angularMomentumTime}. For this specific merger, the
  net increase of the angular momentum is matched by the increase in
  stellar mass, producing only a small change of $j_\star$.

  Around 12.3 Gyr there is another flyby of a galaxy which was thus
  far unable to form stars. When this galaxy enters the dense
  environment of the main galaxy it starts to form stars. After
  passing by the main galaxy, it keeps forming stars resulting in an
  increase of $j_{\star}$, due to more off-center star formation. 

  Mergers involving small halos incapable of forming stars or of
  triggering a star-formation event when captured influence the
  stellar specific angular momentum in a more indirect way:~their
  angular momentum is absorbed by the main halo and can later be
  transferred to newborn stars.
\end{description}


From Fig. \ref{fig:angularMomentum} we conclude that the merger
simulations follow the observational trend, e.g. it is in line with
the dotted trendline of the elliptical galaxies, which tend
to have a lower specific angular momentum than spiral galaxies at a
given stellar mass. This is a consequence of the fact that, depending
on the orbits of the mergers, their mass ratio, their number, etc.,
the specific angular momentum {\em can} be higher in more massive
galaxies. In other words:~a galaxy that formed through a high orbital
angular momentum, late, almost equal-mass merger will end up with a
high stellar specific angular momentum.

\subsection{Kinematics}
\subsubsection{Anisotropy diagram}
\begin{figure}
 \centering
 \includegraphics[width=\columnwidth]{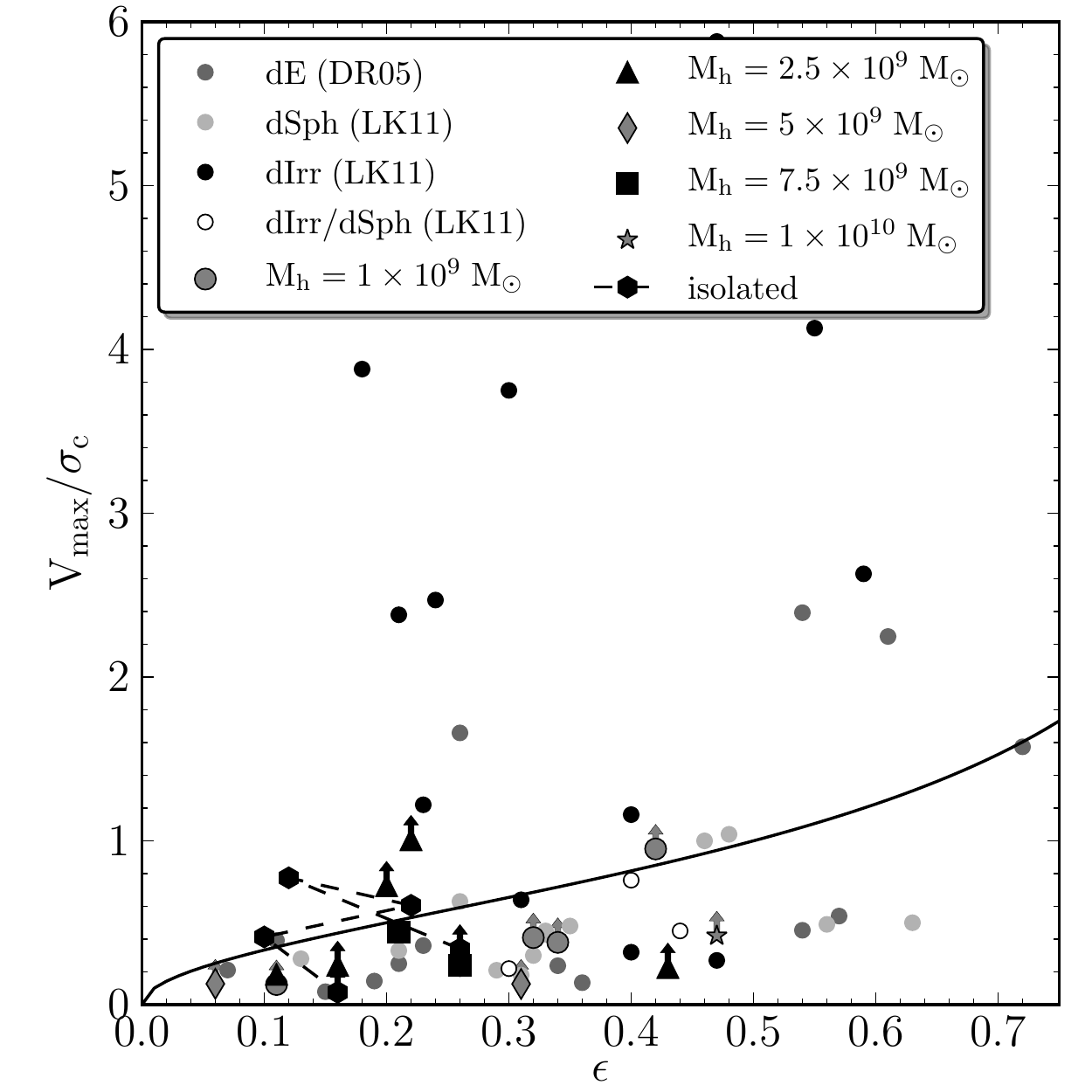}
 \caption{$(V_{max}/\sigma_c)$, the ratio of the maximal rotation
   velocity of the stars and the central velocity dispersion of the
   stars as a function of the ellipticity. The simulations at $z=0$
   are plotted together with dE data from \citet{derijcke05}, and dSph
   and dIrr data from \citet{lokas11}. The black solid line shows the
   $(V/\sigma)$ relation for an oblate isotropic rotator. The arrows
   indicate that the estimation of V$_{max}$ should be considered as a
   lower limit.\label{fig:vsigma}}
\end{figure}

The ratio of the maximum rotational velocity of the stars, $V_{max}$,
and the central velocity dispersion of the stars, $\sigma_c$, is
plotted in Fig. \ref{fig:vsigma} as a function of the ellipticity
$\varepsilon=1-\frac{b}{a}$, with $b$ and $a$ the isophotal minor and major
axis, respectively. To determine the flattening, the simulated galaxy
is first rotated to align the $z$-axis with its rotation axis. Next,
the density is evaluated at the effective radius in the
equatorial plane, this isophote's major axis $a$. Subsequently, the
location along the $z$-axis is determined where the same density is
reached, this isophote's minor axis $b$. From this, the ellipticity of
this isophote immediately follows. The maximum velocity is determined
as the maximum of the least-square fitted function of the following
form to the rotation velocity curve \cite{giovanelli02}:
\begin{equation}
 V(r)=a(1-e^{-r/b})\left(1+c\frac{r}{b}\right).
\end{equation}
In some cases the rotation curve keeps increasing up to the last data
point and the V$_{max}$-value should be considered as a lower
limit. The fitted range that was used depends on the size of the
galaxy and was chosen to be around 5 times the effective radius which
is in agreement with the 'best range' suggested by
\cite{romanowsky12}, between 3 and 6 R$_{e}$, to calculate the
rotation velocity for j$_{\star}$-estimates.
\begin{figure}
 \centering
 \includegraphics[width=\columnwidth]{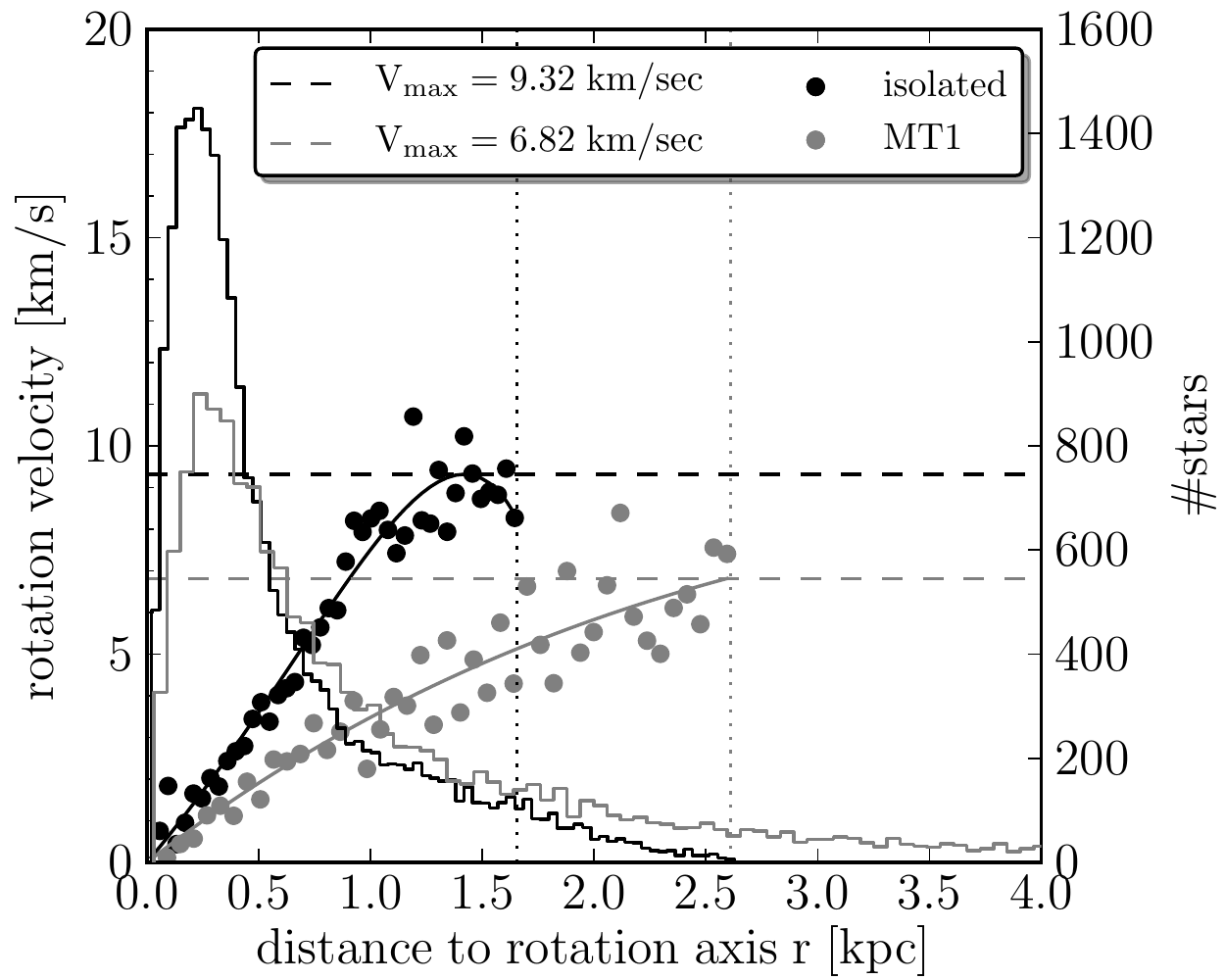}
 \caption{The rotation velocity profile and a histogram of the
   distribution of stars as a function of radius. The dots represent
   the data of respectively the isolated model in black and of MT1 in
   gray and the line is the fit to the data. The dotted line indicates
   the location of 5$R_{e}$ and the dashed line shows the location of
   V$_{max}$.\label{fig:vrot}}
\end{figure}

Fig. \ref{fig:vrot} shows an example of the determination of
V$_{max}$: the rotational velocity profile of an isolated simulation
and of a merger simulation, respectively in black and gray. To each
profile a fit is made and for the isolated simulation this reaches a
maximum while for the merger simulations the fit keep increasing so
V$_{max}$ is taken to be the value at 5$R_{e}$. The location of
5$R_{e}$ is indicated by the dotted line while the value of V$_{max}$
is shown by the dashed line. The histogram shows the distribution of
the stars. In the isolated simulation the stars are more centrally
concentrated and most of the angular momenta is located at large
radii. In the merger simulations, the stellar body extends much
further compared to the isolated simulation.

Most of the simulations are located below the relation for oblate
isotropic rotators defined as $(V/\sigma)_{theo} =
\sqrt{\varepsilon/(1-\varepsilon)}$ and indicated by the black line in
Fig. \ref{fig:vsigma}. This shows that velocity anisotropy plays a
substantial role in stabilizing them. In Table \ref{table_finalprop},
the value $(V/\sigma)^{\star}$ is shown for the simulations, this is
the ratio of $V_{max}/\sigma_{c}$ and the theoretical value for an
isotropic oblate rotator. Hence, a $(V/\sigma)^{\star}$-value of one
corresponds to an isotropic oblate rotator. Most merger simulations
have $(V/\sigma)^{\star}$ values lower then 1. Some of the isolated
simulations have $(V/\sigma)^{\star}$-values much larger than
one. However, their maximum rotation velocity is reached by stars at
the outskirts of the stellar body, beyond $\sim 5$ half-light
radii. Therefore, for these galaxies, no relation between
$V_{max}/\sigma_{c}$ and the stellar body's ellipticity is expected.

\cite{cox06} found that dissipationless and dissipational (with a gas
fraction of 0.4) binary mergers remnants are located in different
locations in the anisotropy diagram, with the former having much lower
$(V/\sigma)^{\star}$-values than the latter. The location of the
merger simulations agrees with the dissipational binary merger
remnants of \cite{cox06}, which could be expected as our simulations
are all gas rich mergers. The merger simulations cover a wider range
in ellipticities, between 0.06 and 0.47, compared to the isolated
models which have ellipticities between 0.10 and 0.26 indicating that
the merger events are efficient in creating flattened
galaxies. However, there is no clear connection between the
characteristics of the merger tree and the final ellipticity. For
example, the merger simulations with a final halo mass of
7.5$\times$10$^{9}$ M$_{\odot}$ have very different merger histories
but have almost identical final ellipticities.

\subsubsection{Shapes - Triaxiality}
\begin{figure}
 \centering \includegraphics[width=\columnwidth]{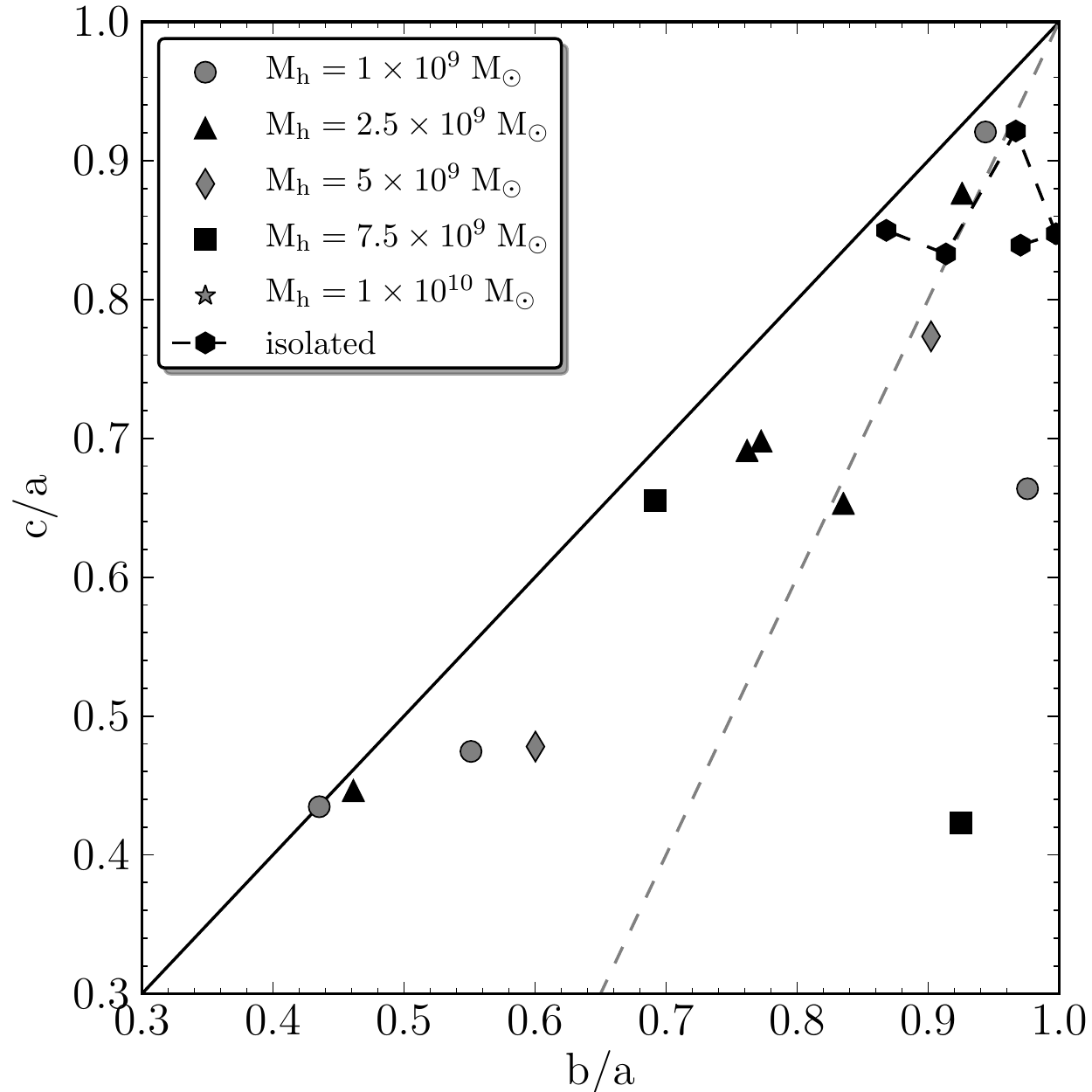}
 \caption{The shape diagram where the axis ratio b/a is plotted as a
   function of $c/a$ with $c<b<a$. The black line corresponds with
   prolate spheroids which for which $b=c$ while oblate spheroids are
   located near the $b/a=1$ line. The dashed line corresponds to the
   models with maximum triaxiality.\label{fig:inertia}}
\end{figure}

In Fig. \ref{fig:inertia}, the shape diagram of the simulated galaxies
is plotted. Each galaxy has first been aligned with the principal axes
of its inertia tensor \citep{franx91, gonzalez05, cox06}. The order of
the three axes is determined from the density profile along each axis,
with $c<b<a$. Then, the two axis ratios $c/a$ and $b/a$ are measured
similarly to the flattening $\varepsilon$ in the previous
paragraph:~the density is evaluated at the effective radius along the
longest axis and those positions along the shortest and intermediate
axes are determined where the same density is reached. Oblate
spheroids have $b/a=1$, which puts them on the right vertical axis of
Fig. \ref{fig:inertia}. Prolate spheroids have $b/a=c/a$ and fall on
the diagonal, marked by a black line, in Fig. \ref{fig:inertia}.

The isolated models are all quite round, with axes ratio above $\sim
0.8$. The models with a merger history, on the other hand, can be
much more flattened, with axis ratios down to 0.4. Moreover, their
shapes can be significantly triaxial. The dashed line in
Fig. \ref{fig:inertia} traces the locus of maximum triaxiality, given
by
\begin{equation}
\frac{c}{a} = 2 \frac{b}{a}-1,
\end{equation}
and many merger models indeed end up close to this line. This is at
least in qualitative agreement with the flattening distribution
analysis of Virgo dwarfs by \citet{bipo95}. These authors find that
the apparent ellipticity distribution of dwarf ellipticals can only be
reproduced by adopting a modest degree of triaxiality, corresponding
to $b/a \sim 0.8-0.9$ (and even smaller $b/a$-values for later type
dwarfs).

\section{Conclusion}
\label{section:conclusion}

We performed a set of simulations of dwarf galaxies with final masses
in the range of 10$^{9}$ M$_{\odot}$ to 10$^{10}$ M$_{\odot}$. We have
shown that simulations based on merger trees constructed by the
\cite{parkinson08} algorithm and using orbital parameters drawn from
the \cite{benson05} velocity distributions are a viable and
time-saving alternative to full-fledged cosmological
simulations. While the simulations presented here do not take into
account all possible effects playing a role in dwarf galaxy evolution,
e.g. they lack a cosmological UV background and external gas removing
processes, they do allow to investigate the effects of the galaxies'
past merger histories on its star-formation history, internal
kinematics, and its dark matter density profile.

The implementation of a hierarchical merger history in the simulations
introduces more variability into the typically periodic SFR of the
isolated simulations. The merger simulations can have short bursts in
their SFH which is likely to be unresolved in the observed SFHs of
dwarfs. The variability of the SFHs of the simulated dwarfs is in
agreement with the complex SFHs that are observed \citep{skillman03,
  monelli10a, monelli10b, weisz11}. The star formation histories of
the galaxies with a merger history show that their stellar mass is
built up more slowly in time compared to the isolated systems.

Mergers can trigger strong star-formation episodes that, through the
concerted feedback of many supernova explosions, can shut down star
formation for up to several gigayears. This impulsive removal of gas
also contributes to the destruction of the central density cusp of the
initial NFW dark matter haloes. Especially in galaxies that grow
through a sequence of minor mergers, each one leading to a short burst
of star formation, the central dark-matter density cusp significantly
flattens over time. The cusp also flattens in isolated galaxies
\citep{cloetosselaer12} but the effect is much more pronounced when
taking mergers into consideration.

Within our merger trees, we consider two main types which have very
different influences on their final properties:~(i) merger trees with
an early massive progenitor that experiences subsequent minor mergers
and (ii) merger trees with many small progenitors that merge only
quite late. The former generally have shallower dark-matter potentials
due to the minor mergers which are more efficient in flattening the
cusp in combination with the larger amount of feedback they experience
as they have larger stellar mass compared to the other type (at a
fixed halo mass). Since there is already a quite massive progenitor
present early on, fewer subsequent mergers are required to build up
the mass of the final galaxy. This gives less opportunity for the
orbital angular momentum of the mergers to cancel, leading to a galaxy
with a higher specific angular momentum.

The latter accumulate their mass more slowly, with generally a major
merger quite late in the simulation. The dark matter density profile
stays more peaked which produces galaxies with smaller half-light
radii and higher stellar surface densities. More mergers are required
to build up the mass of the final galaxy, giving more opportunity to
cancel the orbital angular momentum of the mergers, leading to a
galaxy with a lower specific angular momentum.

All merger-tree simulations have shallower dark-matter potentials than
isolated models of equal mass and in turn lead to galaxies that have
larger effective radii, lower central velocity dispersions, and lower
central surface brightness. They generally overlap with the observed
dwarfs in diagrams where these properties are presented as a function
of luminosity although the trend to become more diffuse with
increasing stellar mass is perhaps stronger than in the observational
data. The $V-I$ and $B-V$ colors are insensitive to the details of the
merger tree. Due to the ongoing star formation, the colors of the
simulated dwarfs are bluer than those of dSphs and are more in
agreement with those of dIrrs.

Except for the least massive merger models, which tend to be too
metal-rich, the merger simulations overlap with the locus of the dSphs
and dIrrs in a metallicity versus luminosity diagram. We show that the
features in the metallicity distribution functions of merger
simulations can also be found in observed dwarfs with similar mean
metallicities, like Fornax, LeoI, Sculptor, and WLM.

We compare the final specific stellar angular momentum of our
simulations with observational data and conclude that they follow the
trend of the observations. The final j$_{\star}$-value of the merger
simulations depends on many variables, such as the orbit of a merger,
its mass ratio, the number of mergers etc. For example, a late major
merger with high orbital angular momentum will result in a galaxy with
a high stellar specific angular momentum. Because of the randomizing
effect of the merger history, j$_{\star}$ can vary by over an order of
magnitude at a given mass.

Most models fall below the locus of the isotropic oblate rotators in
the $v_{max}/\sigma_c$ versus ellipticity diagram. This indicates that
they have significantly anisotropic orbital distributions. This is
corroborated by their place in the shape diagram of $c/a$ versus
$b/a$, with many merger models being strongly triaxial. This is at
least qualitatively in agreement with the observed shapes of dwarf
galaxies.

\section*{Acknowledgements}
We wish to thank the anonymous referee for the many constructive
remarks and questions which greatly improved the contents and
presentation of the paper. Annelies Cloet-Osselaer, Sven De Rijcke and
Bert Vandenbroucke thank the Ghent University Special Research Fund
for financial support.  Joeri Schroyen and Mina Koleva thank the Fund
for Scientific Research - Flanders, Belgium (FWO). Robbert Verbeke
thanks the Interuniversity Attraction Poles Programme initiated by the
Belgian Science Policy Office (IAP P7/08 CHARM).

We thank Volker Springel for
making publicly available the {\sc Gadget-2} simulation code. 

\bibliographystyle{mn2e}
\bibliography{article_bibliography}


\label{lastpage}

\end{document}